\documentclass[sigplan,screen,nonacm]{acmart}
\usepackage{multirow}
\usepackage[htt]{hyphenat}
\usepackage{graphicx}
\usepackage{subcaption}
\usepackage{array}
\usepackage{makecell}
\usepackage{placeins}

\newcolumntype{C}[1]{>{\centering\arraybackslash}p{#1}}
\AtBeginDocument{%
  }

\setcopyright{acmlicensed}
\copyrightyear{2027}
\acmYear{2027}
\acmDOI{XXXXXXX.XXXXXXX}
\acmConference[CGO '27]{International Symposium on Code Generation and Optimization}{February 2027}{Somewhere, Anywhere}
\acmISBN{978-1-4503-XXXX-X/2018/06}

\begin{document}

\thanks{Notice: This manuscript has been authored by UT-Battelle LLC under contract DE-AC05-00OR22725 with the US Department of Energy (DOE). The US government retains and the publisher, by accepting the article for publication, acknowledges that the US government retains a nonexclusive, paid-up, irrevocable, worldwide license to publish or reproduce the published form of this manuscript, or allow others to do so, for US government purposes. DOE will provide public access to these results of federally sponsored research in accordance with the DOE Public Access Plan (\url{https://www.energy.gov/doe-public-access-plan}).}

\title{Opportunistic ZGC: Leveraging Idle Cores for More Effective Concurrent Garbage Collection}

\author{Jacob Malloy}
\email{jmalloy1@utk.edu}
\orcid{1234-5678-9012}
\affiliation{%
  \institution{University of Tennessee}
  \city{Knoxville}
  \state{Tennessee}
  \country{USA}
}

\author{Michael R. Jantz}
\correspondingauthor
\email{mrjantz@utk.edu}
\orcid{1234-5678-9012}
\affiliation{%
  \institution{University of Tennessee}
  \city{Knoxville}
  \state{Tennessee}
  \country{USA}
}

\author{Terry Jones}
\email{trjones@ornl.gov}
\orcid{1234-5678-9012}
\affiliation{%
  \institution{Oak Ridge National Laboratory}
  \city{Oak Ridge}
  \state{Tennessee}
  \country{USA}
}

\renewcommand{\shortauthors}{Malloy, Jantz, and Jones}

\begin{abstract}
Managed language runtimes often provide concurrent garbage collectors so that latency-critical applications with large working sets can keep running while most collection work proceeds in the background.
ZGC is a production-quality, generational, concurrent collector in OpenJDK with sub-millisecond pause times.
While ZGC is designed to run concurrently, frequent and excessive collections with ZGC can still slow the mutators due to synchronization costs and interference with shared computing resources.
Hence, the ZGC scheduler is conservative by default, and in most cases, will grow the heap toward the maximum allowed before scheduling a collection.
While this approach minimizes collection effort, it can be wasteful, or even harmful, if the maximum heap size is not well tuned to the actual working set.

We propose Opportunistic ZGC (OppZGC), a feedback-directed ZGC scheduling policy that constrains the heap \emph{dynamically} and \emph{automatically}, without per-application tuning.
OppZGC identifies periods when CPU cores are underutilized and leverages them for concurrent collection with ZGC.
We describe the design and implementation of OppZGC in OpenJDK's HotSpot Java VM and evaluate it with standard and latency-sensitive benchmarks from DaCapo Chopin and SPECjbb.
OppZGC limits heap usage when there is CPU capacity sufficient for additional collections, and avoids scheduling extra collections when they would substantially degrade performance.
Overall, it reduces maximum heap usage for our DaCapo benchmarks by between 61\% and 90\%, on average, depending on configuration, with minimal impact on throughput and request latency compared to default ZGC.
\end{abstract}


\maketitle

\section{Introduction}
Many of today's most popular programming languages, including Java, Go, Python, and JavaScript, manage memory automatically with garbage collection (GC). 
Runtimes for these languages implement a variety of GC strategies to meet the needs of a broad spectrum of applications and usage scenarios. 
For example, recent versions of the HotSpot Java Virtual Machine (JVM) (from OpenJDK) provide a range of GC options with tradeoffs in throughput and pause times.
ParallelGC, HotSpot's default prior to JDK 9, and G1GC, which is the default since then, are both generational, compacting collectors with multi-threaded, mostly stop-the-world collection.
While these collectors achieve excellent throughput, they can incur long pauses for large heaps, which may be unacceptable for latency-critical applications.

To support such applications, HotSpot introduced ZGC in JDK 11, a fully concurrent collector with sub-millisecond pause times practically independent of the size of the heap~\cite{liden_jep_2018}.
In general, ZGC sacrifices some throughput to enable more fine-grained sharing of computing resources.
It relies on load and store barriers (with colored pointers) to implement incremental marking and relocation of program data objects.

Since allocation and collection can occur simultaneously, ZGC is vulnerable to scenarios where the allocation rate outpaces the collection rate and exhausts the available heap capacity, forcing the mutators to stall.
Even if heap capacity is sufficient to avoid such \emph{allocation stalls}, frequent and excessive collection with ZGC can still harm performance by causing frequent synchronization pauses, increasing the rate of slow paths at barrier instructions, and stealing or interfering with shared computing resources.

To avoid unnecessary work and potential slowdowns, the default ZGC scheduler is relatively conservative.
In most cases, it grows the heap toward the maximum allowed by the JVM instance and defers collection until runtime heuristics predict that delaying further would risk an allocation stall.
While this approach minimizes GC effort, it can be wasteful, or even harmful, if the working set of the application is much smaller than the maximum heap size.
Besides monopolizing memory resources that may be better used by other applications, it can also hurt performance by increasing paging overheads and reducing locality in the heap.

ZGC should thus employ a maximum heap size that is large enough to prevent allocation stalls and mutator interference, but not so large that it wastes memory or harms locality.
Striking the proper balance can be notoriously difficult, and the community currently lacks approaches for tuning ZGC scheduling parameters automatically.
Best practice is empirical testing with JVM tunables such as \texttt{-Xmx} and \texttt{-XX:SoftMaxHeapSize}~\cite{automatedHeapSizingTalk}, which requires manual effort and leaves deployments exposed when production demands diverge from tuned expectations.

We address these gaps with Opportunistic ZGC (OppZGC), a feedback-directed GC scheduling policy that \emph{dynamically} and \emph{automatically} constrains the heap by identifying periods when CPU cores are underutilized and leveraging them for concurrent collection.
It relies on two key observations.
First, many execution scenarios have frequent periods where one or more computing cores are underutilized (e.g., due to over-provisioning or imperfect scaling).
Second, the current ZGC scheduler is often too conservative and wasteful of memory resources.
While more frequent collection can harm mutator performance, a more aggressive scheduler can limit these costs and effectively constrain heap occupancy as long as additional GC cycles: 1) are metered by growth in the heap, and 2) only run on otherwise idle CPU. 

We have built OppZGC by extending the open-source HotSpot JVM in the Java Development Kit (JDK) (jdk-24+2, mainline commit 50bed6c)~\cite{JDKversion}.
The design is straightforward; it extends the ZGC scheduler with two additional rules for inducing collection cycles when sufficient CPU capacity is available.
However, our evaluation, which includes standard benchmarks from DaCapo Chopin v. 23.11-MR2~\cite{dacapo2025asplos} as well as the memory-intensive and latency-sensitive SPECjbb~\cite{specjbb}, demonstrates that this approach effectively constrains memory usage for applications that use ZGC, without increasing the latency of critical operations or reducing application throughput, in almost every case.

This work makes the following important contributions.
\begin{itemize}\itemsep2pt \parskip0pt
\item It describes the design of Opportunistic ZGC: a ZGC scheduling policy that schedules collections when the heap has grown beyond a configurable multiple of its recent occupancy \emph{and} there is enough computing capacity available to complete the cycle.
\begin{sloppypar}
\item It evaluates OppZGC on two x86-64 platforms: an Intel-based server and an AMD-based desktop machine. OppZGC reduces memory usage substantially compared to baseline ZGC with minimal impact on throughput and latency. On the Intel platform, it reduces maximum heap usage for the DaCapo benchmarks by between 61\% and 90\%, on average, depending on the configuration. For SPECjbb, it reduces heap usage by up to 21\%, with no harm to maximum or critical jOPS in large memory configurations.
\end{sloppypar}
\item It evaluates OppZGC with varying amounts of CPU capacity available to the Java process. OppZGC effectively adapts its scheduling policy to CPU-constrained environments. With only two cores available, OppZGC still exhibits similar throughput and latency performance as default ZGC and reduces maximum heap usage for the DaCapo default and large input sets by 63\% and 39\%, respectively.
\end{itemize}

\vspace{-2ex}
\section{Concurrent Collection with ZGC}
This section describes the design and implementation of ZGC from JDK~24, which is the basis for OppZGC, as well as current options for tuning application heap sizes with ZGC.


\subsection{ZGC Design and Implementation}
ZGC is a mark-evacuate concurrent collector built for parallel architectures.
Almost all of its collection work, including marking and relocation, is multi-threaded and performed while mutators are running.
ZGC's heap is divided into regions, called ZPages, which are based on size classes.
New objects are bump-pointer allocated into the ZPage of the appropriate size class. 
ZGC periodically defragments the heap by copying surviving objects from sparse ZPages, which are then freed.
This design allows ZGC to control collection costs because it can choose to relocate objects on only a fraction or none of the ZPages, depending on need.
Generational ZGC~\cite{stefan_karlsson_jep_2021}, introduced in JDK~21 and now the only mode available in JDK~24, additionally partitions ZPages based on the age of their data and performs separate cycles for minor (young-only) and major (young and old) collections.

\begin{table*}
   \small
   \begin{tabular}{|p{1.7cm}|p{3.4cm}|p{11.2cm}|}
   \hline
   \multicolumn{1}{|c|}{\textbf{Type}} &
   \multicolumn{1}{|c|}{\textbf{Name}} &
   \multicolumn{1}{|c|}{\textbf{Fires When}}\\
   \hline
   \multirow{5}{1.7cm}{\centering \textbf{Major GC\\Rules}} &
   \texttt{major\_timer} &
   Time since last major GC exceeds \texttt{ZCollectionIntervalMajor}. Disabled by default.\\
   \cline{2-3}
   &
   \texttt{major\_warmup} &
   Heap usage crosses threshold percentages of the soft maximum. Startup rule only.\\
   \cline{2-3}
   & 
   \texttt{major\_proactive}&
   Enough time or growth has passed that the throughput cost of major GC is acceptable.\\
   \cline{2-3}
   &
   \multirow{2}{3.4cm}{\texttt{major\_allocation\_rate}} &
   An already-warranted minor collection should be upgraded to a major collection, typically because the estimated cost efficiency of major collection is higher than minor collection.\\
   \hline
   \multirow{3}{1.7cm}{\centering \textbf{Minor GC\\Rules}} &
   \texttt{minor\_timer} &
   Time since last minor GC exceeds \texttt{ZCollectionIntervalMinor}. Disabled by default.\\
   \cline{2-3}
   &
   \texttt{minor\_allocation\_rate} &
   Predicted allocation rate implies heap will be exhausted before a young collection finishes.\\
   \cline{2-3}
   &
   \texttt{minor\_high\_usage} &
   Free memory drops below 5\% of soft max. Backstop for cases with low allocation rate.\\
   \hline
   \end{tabular} 
   \caption{ZDirector rules for initiating major and minor GC.}
   \label{tab:zgc_rules}
\end{table*}
\subsubsection{Colored Pointers and Barrier Instructions}
To ensure that the concurrent mutator and GC threads always see valid pointers, ZGC uses colored pointers and barrier instructions.
Object references in the ZGC heap are always 64 bits: address bits plus meta bits describing the pointer's current ``color''.
ZGC maintains global status bits, which mutators check against a pointer's color to determine if the pointer is ``good'' (i.e., valid and usable with no extra work required) in the current execution context, and updates them as the cycle advances.
In generational ZGC, these bits can also distinguish ``store-good'' from ``load-good'' pointers, since a pointer that is valid to dereference may still require GC work before it can be overwritten.

Since collector threads can relocate program objects while the mutator threads are running, ZGC inserts load barriers at pointer read instructions to avoid using or following stale pointers.
If the color of the pointer being read is load-good, the barrier fast path simply strips the pointer color and allows the mutator to use the colorless pointer.
Otherwise, the barrier slow path determines if the object is (or is about to be) relocated, and if so, finds (or decides) the new address of the object.
The pointer is then \emph{healed} by writing the updated, good-colored pointer back to the field from which it was loaded.
In this way, subsequent loads of the same field will use the fast path until the ZGC cycle updates the status bits.

ZGC employs store barriers at pointer writes to add color to colorless pointers held on the stack or in registers before they are written to the heap.
Store barriers serve two other purposes: 1) maintaining the \emph{remembered set} (the set of old generation fields that may contain references to young objects), which is necessary to avoid scanning the old generation during minor collections, and 2) implementing its snapshot-at-the-beginning (SATB) approach for marking, which ensures a consistent view of the heap if a mutator overwrites a pointer during marking~\cite{yuasa1990}.
ZGC fuses the checks for both cases into a single fast-path test at each pointer write.
If either check fails, the slow path performs the required work and recolors the field. 

\begin{figure}
    \centering
    \includegraphics[width=0.9\linewidth]{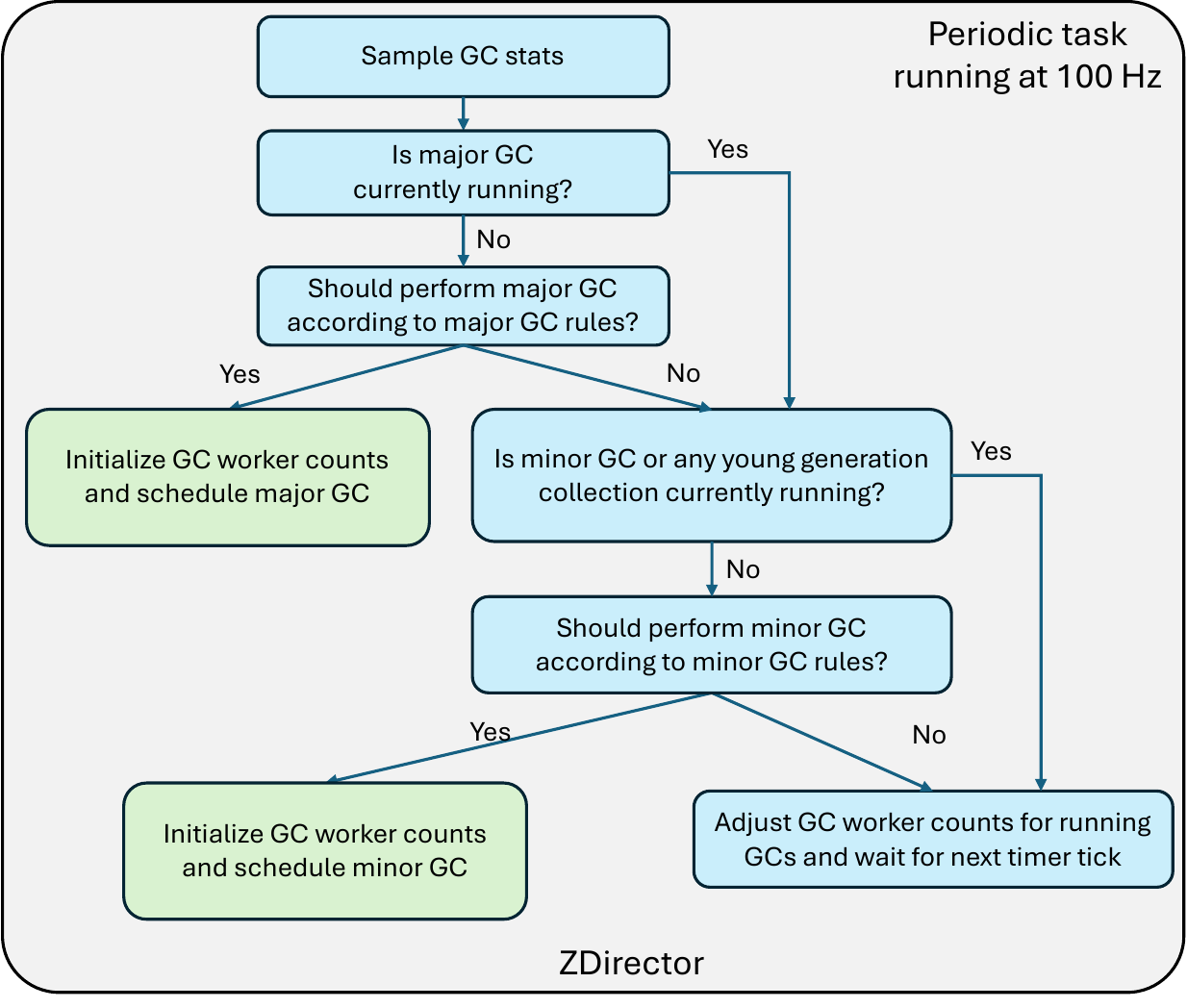}
    \caption{ZDirector decisions and control flow. For simpler presentation, this figure omits cases where an already-warranted minor GC may be upgraded to major GC.}
    \Description{Flowchart of the ZDirector periodic task.} 
    \label{fig:zdirector}
\end{figure}
\subsubsection{Scheduling ZGC Collections}
While certain events, such as allocation stalls and explicit GC requests from the application, can initiate ZGC collections, minor and major GCs are primarily invoked from an asynchronous periodic task, called ZDirector.
Figure~\ref{fig:zdirector} diagrams the control flow and major decisions made by ZDirector.

ZDirector operates at a fixed frequency of 100\,Hz.
At each timer tick, it collects statistics on the allocation rate (averaged and smoothed over recent intervals), current heap usage, and per-generation accounting such as timing stats and bytes reclaimed by previous collections. 
It then uses this information to check a series of rules that determine whether or not to initiate a GC cycle.
If a major collection is not currently running, it checks the rules for running a major collection.
Then, if no major GC is scheduled or running, it checks if minor GC, or the young cycle of a major GC, is already running, and if not, it checks the rules for minor collections.
If either path fires, ZDirector sets the initial number of GC workers based on its sampled statistics and schedules a minor or major GC to start in a separate thread.
Otherwise, it may still adjust the worker count for active collection cycles before sleeping until the next timer tick.

Table~\ref{tab:zgc_rules} presents the ZDirector rules.
The timer rules are unused by default and \texttt{major\_warmup} fires at most three times during startup.
The other rules are conservative by design.
\texttt{major\_proactive} waits until the heap grows by at least 10\% or 5 minutes have passed since the last major GC, and assumes 50\% mutator throughput reduction during collection.
\texttt{minor\_high\_usage} only fires when free memory is below 5\% of the soft maximum allowed and mainly serves as a backstop for when the allocation rate rules do not fire because new allocations have slowed to a trickle.

In practice, most collections are triggered by the \emph{minor\_allocation\_rate} rule.
\emph{major\_allocation\_rate} is checked only after a minor rule fires, and may upgrade the impending minor GC to a major GC.
Since these rules defer collection until new allocations are expected to exhaust free memory within a short time, the ultimate effect of this strategy is to grow the heap until it is near the soft maximum allowed.

\subsubsection{Young and Old Collection Cycles}
Generational ZGC has two types of collection cycles: minor GCs perform only the young collection cycle, while major GCs perform the young collection followed by the old.
Each cycle consists of a small number of sub-millisecond STW pauses and several concurrent phases.
Appendix~\ref{sec:zgc_cycles} describes the cycles in more detail.
Yang and Wrigstad~\cite{yang_deep_2022} provide a comprehensive description of the underlying collection mechanisms, though for an earlier, single-generation version of ZGC (JDK~15).

\begin{figure*}
  \centering
  \begin{subfigure}{0.24\textwidth}
    \centering
    \includegraphics[width=\linewidth]{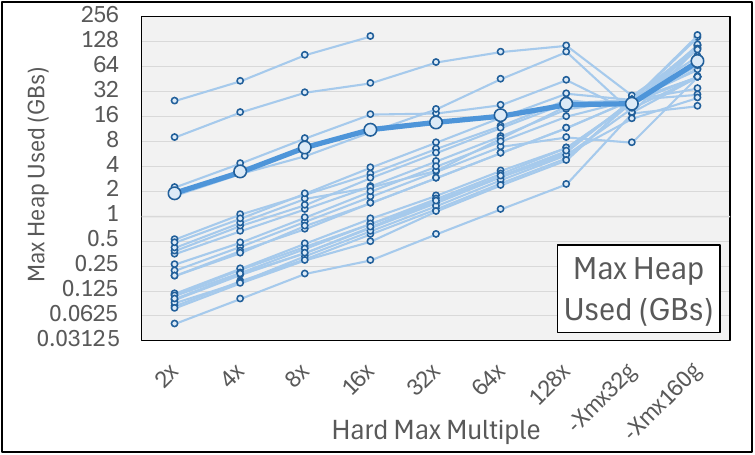}
    \caption{Max heap used (GB, log scale).}
    \label{fig:hard_max_used}
  \end{subfigure}
  \hfill
  \begin{subfigure}{0.24\textwidth}
    \centering
    \includegraphics[width=\linewidth]{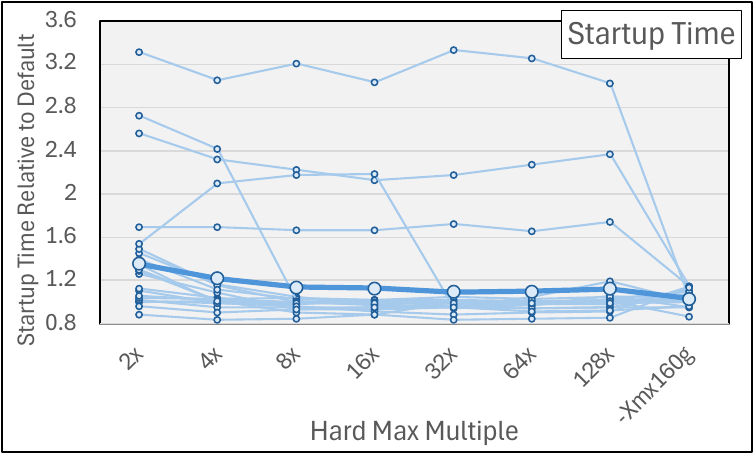}
    \caption{Startup time.}
    \label{fig:hard_startup}
  \end{subfigure}
  \hfill
  \begin{subfigure}{0.24\textwidth}
    \centering
    \includegraphics[width=\linewidth]{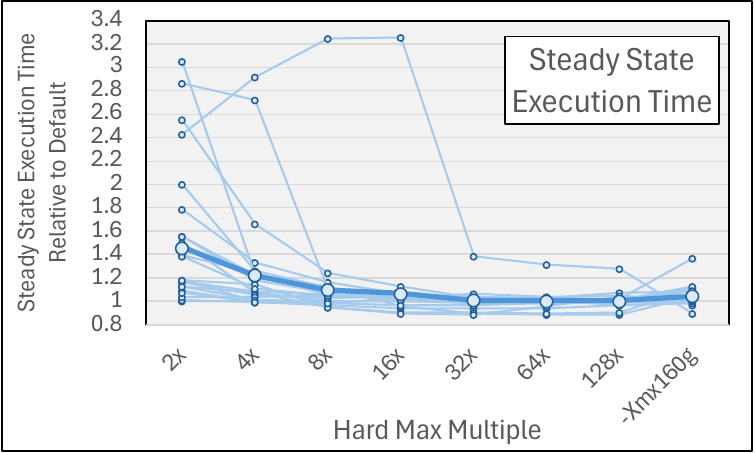}
    \caption{Steady-state execution time.}
    \label{fig:hard_steady_state}
  \end{subfigure}
  \hfill
  \begin{subfigure}{0.24\textwidth}
    \centering
    \includegraphics[width=\linewidth]{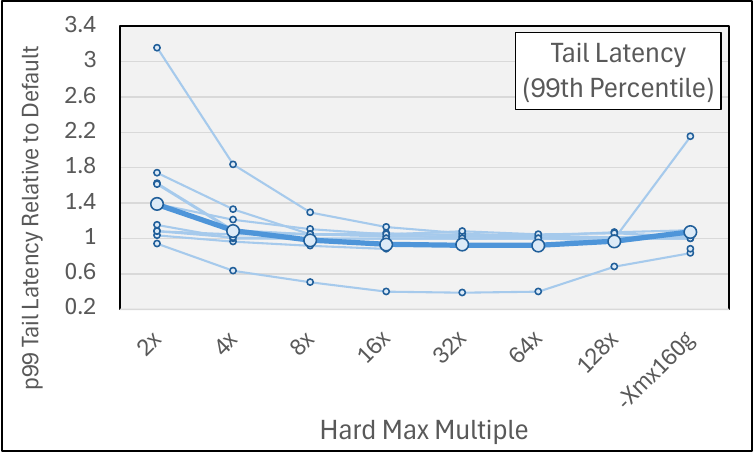}
    \caption{Tail latency (99th percentile).}
    \label{fig:hard_latency}
  \end{subfigure}
  \caption{DaCapo benchmarks with hard max (-Xmx) set to different multiples of the minimum heap size.}
  \Description{Memory usage and performance for the DaCapo benchmarks with hard maximum set to different multiples of the minimum heap size.}
  \label{fig:hard_max}
\end{figure*}

\begin{figure*}
  \centering
  \begin{subfigure}{0.24\textwidth}
    \centering
    \includegraphics[width=\linewidth]{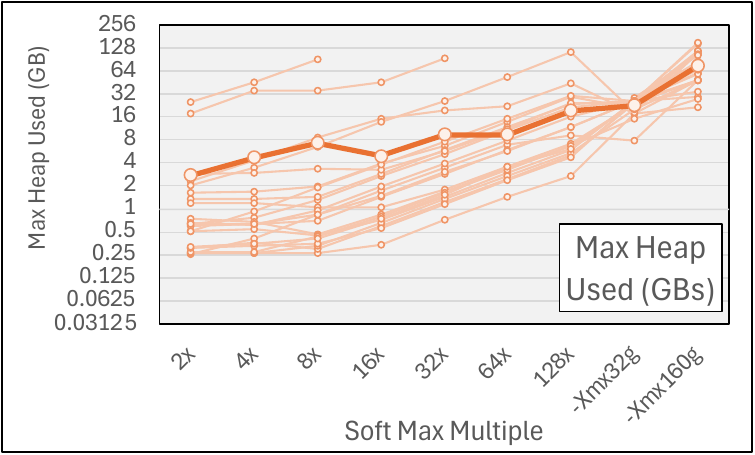}
    \caption{Max heap used (GB, log scale).}
    \label{fig:soft_max_used}
  \end{subfigure}
  \hfill
  \begin{subfigure}{0.24\textwidth}
    \centering
    \includegraphics[width=\linewidth]{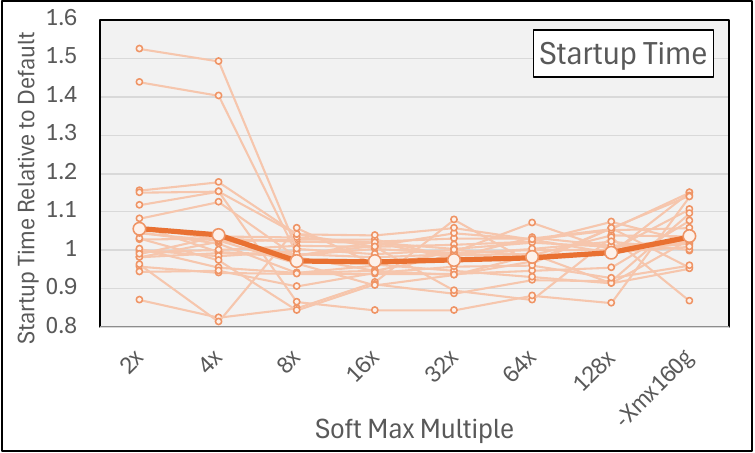}
    \caption{Startup time.}
    \label{fig:soft_startup}
  \end{subfigure}
  \hfill
  \begin{subfigure}{0.24\textwidth}
    \centering
    \includegraphics[width=\linewidth]{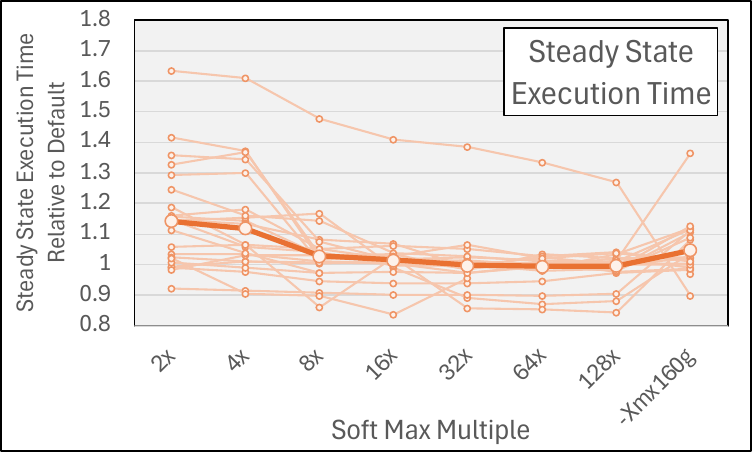}
    \caption{Steady-state execution time.}
    \label{fig:soft_steady_state}
  \end{subfigure}
  \hfill
  \begin{subfigure}{0.24\textwidth}
    \centering
    \includegraphics[width=\linewidth]{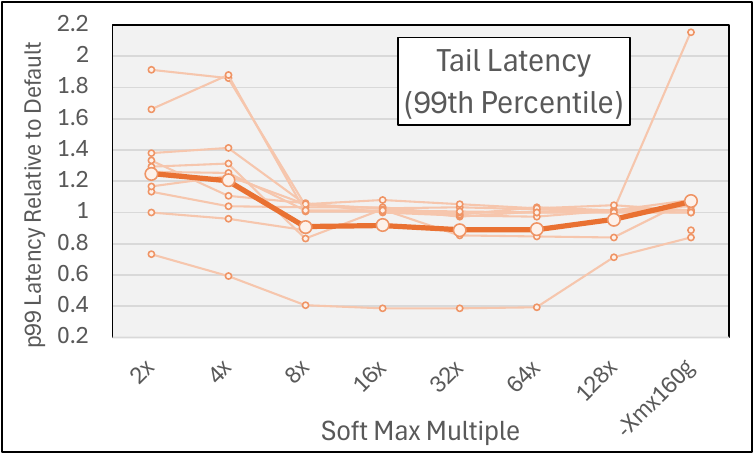}
    \caption{Tail latency (99th percentile).}
    \label{fig:soft_tail}
  \end{subfigure}
  \caption{DaCapo benchmarks with soft max (-XX:SoftMaxHeapSize) set to different multiples of the minimum heap size.}
  \Description{Memory usage and performance for the DaCapo benchmarks with soft maximum set to different multiples of the minimum heap size.}
  \label{fig:soft_max}
\end{figure*}
\subsection{Controlling the Frequency of ZGC}
\label{sec:hardmax_softmax}
Beyond explicit calls (e.g., \texttt{System.gc()}) and the timer rules above, users primarily control ZGC frequency by limiting the size of the heap.
Java provides two command line options for this purpose: \texttt{-Xmx}, which is a hard limit, and \texttt{-XX:SoftMaxHeapSize}, which is a soft limit that ZGC strives to stay within.
The hard max defaults to 25\% of physical memory, up to a maximum of 32 GB, and the soft max defaults to the hard max.
ZDirector proactively invokes collections to try to keep the heap occupancy under the soft max.
If a new allocation would breach the hard max, the mutators will stall until a collection has freed enough space (or, if none can be freed, the JVM reports out of memory).  

To illustrate the effect of these options with ZGC, we measured the minimum heap sizes of each of our DaCapo benchmarks with default ZGC (both default and large inputs, using the methodology described in \cite{dacapo2025asplos}), and ran each workload in isolation on our Intel platform with hard max or soft max set to one of a series of multiples of its minimum size.\footnote{The minimum heap sizes we measured are listed in Table~\ref{tab:dacapo} of Appendix~\ref{sec:dacapo_table}.}
We compare against default configurations with the hard maximum set to 32 GB (the Java-selected default on our server platform) and with the hard maximum set to 160 GB (most of the 192 GB available).
The soft max experiments all use a hard max of 192 GB, which is never reached.

Figures~\ref{fig:hard_max} and \ref{fig:soft_max} show the maximum heap usage and performance, in terms of startup time, steady-state execution time, and, for the ten benchmarks that report latency statistics, the p99 tail latency, for each hard max and soft max configuration.
In each figure, each line represents a single benchmark, and the thicker, darker line shows the average.
Aside from the heap usage results, all results are shown relative to the baseline ZGC configuration with -Xmx set to 32 GB.
Some lines are incomplete because some benchmarks crash if the maximum heap size is larger than the available memory.

Observe that both hard max and soft max can effectively constrain heap memory utilization.
At lower multiples, hard max uses significantly less memory than soft max, but these savings come at a steep performance price.
For example, with the limits set at $2\times$ the minimum, hard max is about 30\% slower, on average, in both startup and steady-state execution time and its average tail latency is about 10\% worse than soft max.
These degradations are mainly the result of frequent \emph{allocation stalls}, which pause the mutators until a collection can free up enough space to satisfy a new object.

While restricting the soft maximum also enables substantial memory savings, this approach exhibits much less performance downside.
The most aggressive soft max configuration ($2\times$) uses only 12\% of the heap memory used by the default \texttt{-Xmx32g} configuration with startup and steady-state execution times only 6\% and 13\% worse, on average.
Moreover, the $16\times$ soft max configuration reduces memory usage by more than 77\% with almost no performance downside.

\begin{table*}
   \small
   \begin{tabular}{|p{2cm}|p{2.3cm}|p{5.9cm}|p{6.3cm}|}
   \hline
   \multicolumn{1}{|c|}{\textbf{Metric}} &
   \multicolumn{1}{c|}{\textbf{Description}} &
   \multicolumn{1}{c|}{\textbf{How It Is Computed}} &
   \multicolumn{1}{c|}{\textbf{Example}}\\
   \hline
   \multicolumn{1}{|c|}{\multirow{3}{*}{idle\_cores}}
   &\multirow{3}{2.3cm}{\centering\arraybackslash Currently idle CPU capacity (in cores)}
   & \multirow{3}{5.9cm}{\shortstack{Active processor count $-$ \\CPU consumed over the previous $\delta$ time units}}
   & If $\delta=100$ ms and if 10 of 12 cores are utilized at 90\% over the previous $100$ ms, then:\newline
   \centerline{$\text{idle\_cores} = 12 - (10 \times 0.9) = 3.$}\\
   \hline 
   \multicolumn{1}{|c|}{\multirow{3}{*}{gc\_duration}}
   & \multirow{3}{2.3cm}{\centering\arraybackslash Wall clock duration of a GC cycle}
   & \multicolumn{1}{c|}{\multirow{3}{*}{$\text{Serial GC time} + \frac{\text{Parallelizable GC time}}{\text{GC Worker Count}}$}}
   & If serial GC time = 100 ms and parallelizable GC time = 400 ms, and there are 4 GC workers, then: \newline \centerline{$\text{gc\_duration} = 100\,\text{ms} + \frac{400\,\text{ms}}{4} = 200\,\text{ms}.$}\\
   \hline
   \multicolumn{1}{|c|}{\multirow{3}{*}{gc\_cpu\_cost}}
   & \multirow{3}{2.3cm}{\centering\arraybackslash Total CPU cost (in CPU time) of a GC cycle}
   & \multicolumn{1}{c|}{\multirow{3}{*}{$\text{Serial GC time} + \text{Parallelizable GC time}$}}
   & If serial GC time = 100 ms and parallelizable GC time = 400 ms, then:\newline \centerline{$\text{gc\_cpu\_cost} = 100\,\text{ms} + 400\,\text{ms} = 500\,\text{ms}.$}\\
   \hline
   \end{tabular}
   \caption{Metrics used to implement major GC and minor GC rules for OppZGC. Assuming sufficient growth in heap occupancy, the example will trigger GC because it passes Condition \ref{eq:oppzgc_condition} (i.e., $3 \times 200\,\text{ms} = 600\,\text{ms} \ge 500\,\text{ms}$).}
   \label{tab:oppzgc_metrics}
\end{table*}
Furthermore, some workloads actually perform better with a tighter hard or soft maximum than with a mostly unrestricted heap.
With soft max set to $8\times$ the minimum, \texttt{spring} with its default input reduces startup and steady-state time by 13\% and 15\% and tail latency by 17\%.
These gains are primarily driven by reduced paging costs.
For \texttt{spring} specifically, the default configuration incurs $19\times$ as many page faults and 21\% more DTLB misses.

These trends also hold under varying CPU constraints.
Repeating the soft max sweep with the cores restricted to 8, 4, or 2 CPUs (see Figure~\ref{fig:core_soft_max} in Appendix~\ref{sec:core_constraints}) yields broadly similar relative performance at most multiples, but the most aggressive configurations degrade further.
With only two cores and soft max at $2\times$, startup and steady-state slowdowns grow to 15\% and 20\%, and tail latency degrades by over 70\%.

While tuning the hard or soft maximum can yield substantial benefits, this approach is infeasible in many real-world scenarios.
For many applications, heap usage is either poorly understood or unpredictable in a way that makes it difficult to select an appropriate maximum.
Furthermore, since GC workers can steal resources from mutators, GC effort must also be balanced against limits on CPU time, increased barrier slow path rate, and potential interference in shared caches.
For these reasons, HotSpot developers recommend that ZGC users choose heap limits that are large enough to avoid potential performance problems, even if the chosen size is likely to be wasteful of memory capacity~\cite{oracle_tuning,osterlund_jep_2026}.

\vspace{-2ex}
\section{Opportunistic ZGC: Leveraging Idle Cores for More Effective Concurrent GC}
\label{sec:design}
The previous section shows there is a need for an automated approach that effectively constrains the ZGC heap, without slowing the mutators with interference from ZGC workers.
OppZGC fills this gap with a GC scheduling policy that runs GC workers in otherwise idle CPU capacity.

\subsection{OppZGC Collection Rules}
OppZGC extends ZDirector with two new rules, \texttt{major\_opportunistic} and \texttt{minor\_opportunistic}, that fire when the runtime estimates there is enough idle CPU to complete the cycle.
Each rule is checked only after the other rules for that GC type have all been checked and failed to fire.

Running GC in every possible window, even if there is idle CPU capacity, could reduce mutator throughput by increasing the frequency of STW pauses and barrier slow paths.
Hence, each rule first checks if the relevant generation's heap occupancy has grown enough since the previous opportunistic collection to justify another cycle.
Specifically, growth in heap occupancy since that collection, normalized by the amount of garbage it reclaimed, must exceed a configurable \emph{growth ratio} $\alpha$.
$\alpha$ values below 1.0 use free computing capacity to exert downward pressure on the heap, while $\alpha$ values at or slightly above 1.0 cap heap growth to be at or near what was recently reclaimed.
During warmup, this check is skipped if there are no previous opportunistic collections.

If the heap has grown beyond the $\alpha$-based threshold, the OppZGC rules then compute an estimate to predict if there is currently enough idle CPU capacity to run a major or minor GC cycle.
For this estimate, OppZGC employs three runtime-derived metrics, which are shown in Table~\ref{tab:oppzgc_metrics}.

ZGC already maintains running averages of the serial and parallelizable times of previous cycles, and OppZGC uses these averages to estimate gc\_duration and gc\_cpu\_cost.
For idle\_cores, it measures the CPU capacity (in cores) used over the past $\delta$ time units, where $\delta$ is a configurable multiple of the ZDirector timer tick, and subtracts that from the total cores available.
We considered alternative strategies for estimating CPU utilization, including using a moving average to smooth bursty periods and discounting CPU usage by GC workers, but found no benefit to these approaches in initial testing.

After computing the necessary metrics, the OppZGC rules each check the condition:
\begin{equation}
\text{idle\_cores} \times \text{gc\_duration} \ge \text{gc\_cpu\_cost}
\label{eq:oppzgc_condition}
\end{equation}
If the condition is met, then the rule initiates the appropriate GC cycle.
Otherwise, no action is taken and ZDirector continues without invoking a GC cycle as shown in Figure~\ref{fig:zdirector}.

Since past CPU utilization is not always predictive of future behavior, these rules may still decrease mutators' access to the CPU and harm application performance.
However, even in these cases, the costs are still limited by two factors: 1) ZGC collections are mostly concurrent with only sub-millisecond pauses, and 2) OppZGC is self-modulating; that is, if the combined utilization of mutators and GC workers saturates the CPU, no new collections will be performed until there is idle capacity available (or another rule fires).

\begin{figure*}[t]
  \centering
  \begin{subfigure}{0.49\textwidth}
    \centering
    \includegraphics[width=\linewidth]{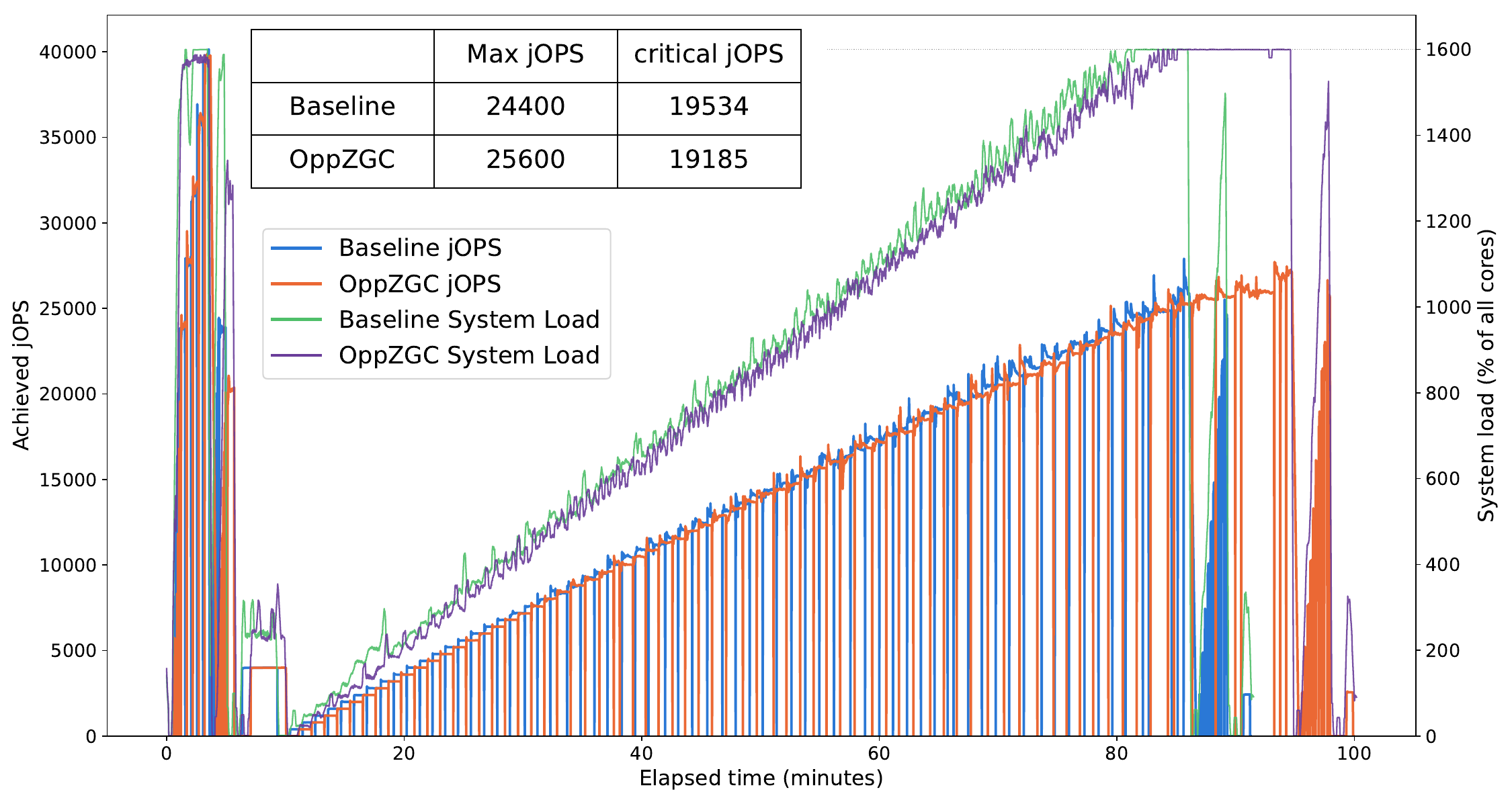}
    \caption{System Load (CPU Utilization) and Performance in jOPS}
    \label{fig:specjbb_system_load}
  \end{subfigure}
  \hfill
  \begin{subfigure}{0.49\textwidth}
    \centering
    \includegraphics[width=\linewidth]{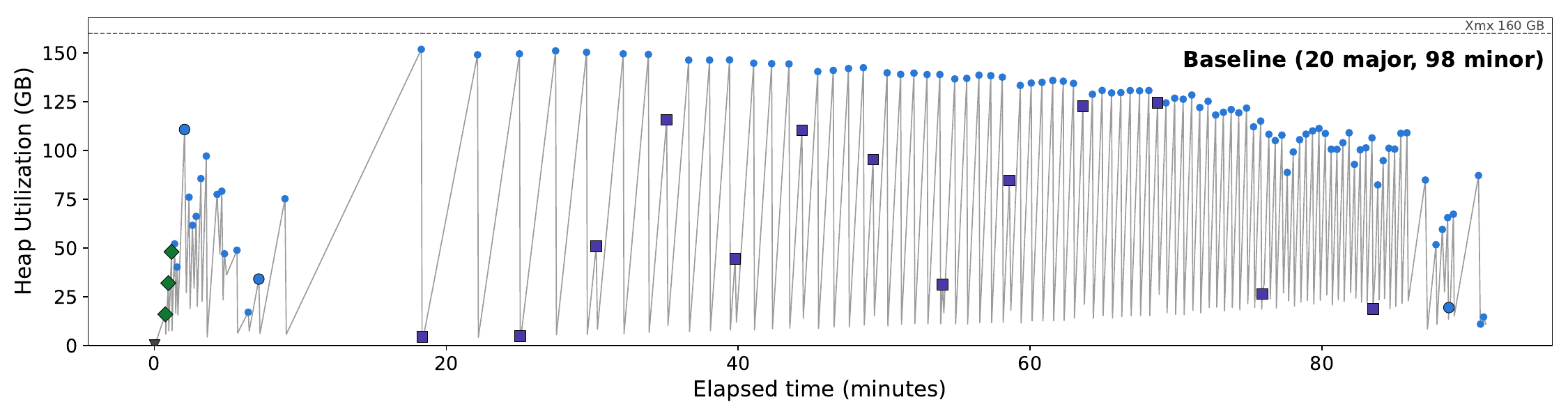}
    \\[1ex]
    \includegraphics[width=\linewidth]{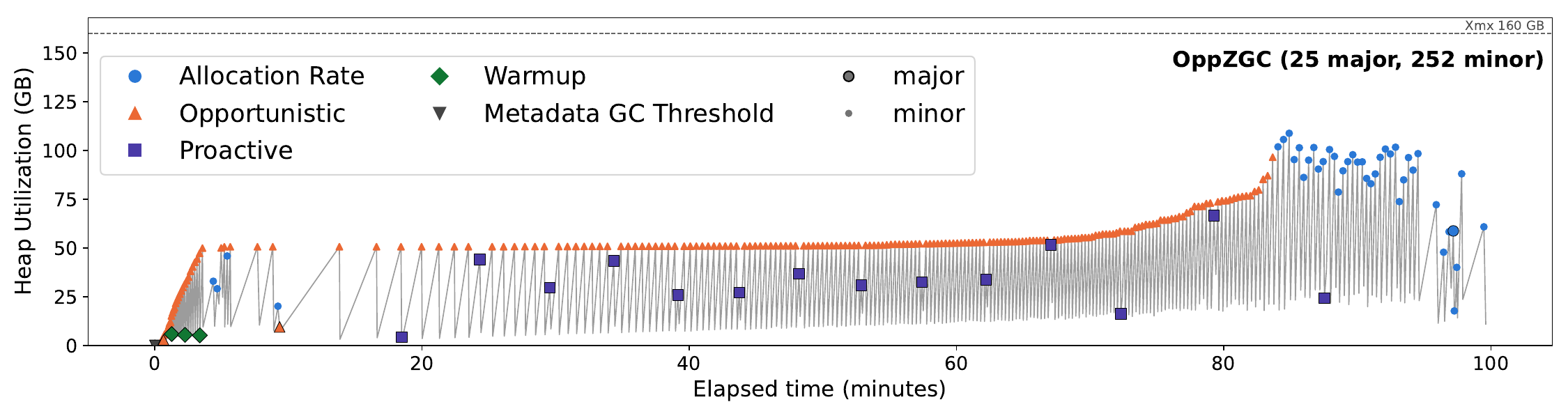}
    \caption{Collection Cycles with Baseline ZGC and OppZGC}
    \label{fig:specjbb_collections_default}
  \end{subfigure}
  \caption{Example Execution of SPECjbb with OppZGC.}
  \Description{Graphs of system load and collection cycles over time with baseline ZGC and OppZGC.}
  \label{fig:specjbb_example}
\end{figure*}
\subsubsection{Implementation Notes and Details}
Our implementation employs the \texttt{getrusage} system call, which reports the CPU utilization of the calling process~\cite{getrusage}.\footnote{\texttt{getrusage} supports $\mu$s resolution, but the actual resolution is platform-dependent. Both of our experimental platforms support $\mu$s resolution.}
This approach is only appropriate for scenarios where the JVM runs on an otherwise idle machine.
\texttt{/proc/schedstat}, for true system-wide CPU utilization, or the cgroup filesystem's \texttt{cpu.stat}, for per-cgroup CPU utilization, can be used for multi-process and multi-tenant usage scenarios.
OppZGC does not scale the number of GC workers based on the idle CPU capacity, but rather uses the GC worker counts recommended by ZDirector, which are based on the size of the heap.
During warmup, before estimates of gc\_duration and gc\_cpu\_cost exist, the OppZGC rules check if there is at least one idle core available, rather than Condition~\ref{eq:oppzgc_condition}.
To avoid initiating multiple collections from stale information, OppZGC starts at most one GC cycle per $\delta$ period.

\subsection{Example execution with OppZGC}
\begin{sloppypar}
Figure~\ref{fig:specjbb_example} shows a detailed timeline of the total system load, memory usage, and collections during execution of the SPECjbb benchmark with two configurations: one with default ZGC and another with OppZGC enabled. 
For this comparison, both configurations set the hard maximum heap limit (\texttt{-Xmx}) to 160 GB, and the OppZGC configuration uses $\alpha=1.0$ and $\delta=20\,\mathrm{ms}$.
A detailed description of the SPECjbb workload is provided in Section~\ref{sec:specjbb}.
\end{sloppypar}

Observe that, during the initial high-intensity HBIR phase, there is still enough computing capacity available for frequent opportunistic collections to keep the heap occupancy below 60 GB.
As system load ramps up during the RT-curve building phase, the baseline configuration quickly exhausts most of the remaining heap capacity, while OppZGC slows growth in the heap with additional minor collections.
When the workload fully saturates the CPU, opportunistic collections stop because there is no longer CPU capacity available for extra collections.
Other ZGC collection rules still fire during this phase, matching the behavior of baseline ZGC.

In this example, OppZGC causes relatively few major GCs.
Since most data in SPECjbb dies before it is tenured, the old generation reaches its largest point early in the run, and the checks for additional growth in the old generation heap fail after this early point.
In both configurations, most major cycles are caused by \texttt{major\_proactive}.
While the proactive rule has a similar goal to OppZGC, its heuristic is overly conservative because it assumes a 50\% throughput drop during collection and does not account for idle computing capacity.
Hence, the proactive rule alone does not limit heap usage nearly as much as OppZGC.

Overall, OppZGC constrains heap occupancy with only marginal impact on overall system load and performance.
The maximum heap occupancy is reduced by about 21\% compared to default ZGC.
At the same time, SPECjbb peak throughput (max jOPS) and latency-constrained throughput (critical jOPS) are essentially the same in both configurations.

\vspace{-2ex}
\section{Experimental Setup}
\subsection{Experimental Platforms and Configuration}
\subsubsection{Testbed Hardware}
Our primary evaluation platform contains a single Intel\textsuperscript{\textregistered{}} Xeon\textsuperscript{\textregistered{}} Gold 6246R CPU (codenamed Cascade Lake) with 16 physical compute cores and hyperthreading disabled.
The cores all run a 3.4~GHz clock and share a 35.75~MB L3 cache.
The processor includes a memory controller that services six channels, each of which is connected to one 32~GB, 2933~MT/s, DDR4 DIMM giving 192~GB of DDR4 SDRAM. 
To confirm the robustness of our results, we also evaluate OppZGC on an AMD Linux-x86-64 platform.
A description of this platform, its configuration, and its results is available in Appendix~\ref{sec:amd_results}.

\subsubsection{System and Runtime Software}
Both testbeds run Debian~11 with Linux~V6.1.27 as the base kernel.
The base JVM is the open-source HotSpot JVM from OpenJDK~24 (build \texttt{jdk-24+2})~\cite{JDKversion}.
Our evaluation uses the server optimized build of HotSpot, which we compiled with -O3 using the GNU Compiler Collection (GCC)~10.2.1.
All of our experiments enable ZGC with \texttt{-XX:+UseZGC} and log collection decisions to a file on disk using HotSpot's built-in logging facilities.
The DaCapo benchmarks always invoke \texttt{System.gc()} between iterations.
To evaluate OppZGC without influence from explicit collections, our experiments also disable explicit collections with \texttt{-XX:+DisableExplicitGC}. 

\subsubsection{Common Experimental Configuration}
Each experiment runs the benchmark on an otherwise idle machine, and all results report the mean of five experimental runs.
To estimate and report variability in our results, we compute 95\% confidence intervals (CIs) for the difference between the means of the experimental and default configurations, using the unequal-variance (Welch) formulation with Welch–Satterthwaite degrees of freedom, as described in Georges et al.~\cite{georges2007oopsla}.
Figures that report results for individual benchmarks plot these intervals as error bars around the sample means.\footnote{Figures~\ref{fig:hard_max} and \ref{fig:soft_max} omit these intervals to avoid cluttering the graphs.}
Some figures report average results over a set of benchmarks.
Average results relative to a baseline configuration present the geometric mean and non-relative averages present the arithmetic mean.


\subsection{Benchmarks}
We evaluate OppZGC with a selection of benchmarks from the DaCapo Chopin suite (v. 23.11-MR2)~\cite{dacapo2025asplos} and the compute- and memory-intensive SPECjbb 2015 (v. 1.03)~\cite{specjbb}.
\begin{table*}[t]
\small
\centering
\setlength{\tabcolsep}{3pt}
\begin{tabular}{|c|c|r|r|r|r|r|r|r|r|r|r|r|r|r|r|r|r|}
\hline
\multirow{3}{*}{Benchmark} &
\multirow{3}{*}{Input} &
\multicolumn{4}{c|}{Baseline ZGC} &
\multicolumn{6}{c|}{OppZGC, $\alpha=1.0$, $\delta=20\,\mathrm{ms}$} &
\multicolumn{6}{c|}{OppZGC, $\alpha=0.9$, $\delta=20\,\mathrm{ms}$}\\
\cline{3-18}
 & & \multicolumn{1}{c|}{Minor} &
     \multicolumn{1}{c|}{Major} &
     \multicolumn{1}{c|}{\multirow{2}{*}{MB/s}} &
     \multicolumn{1}{c|}{\multirow{2}{*}{GCt (s)}} &
     \multicolumn{2}{c|}{Minor} &
     \multicolumn{2}{c|}{Major} &
     \multicolumn{1}{c|}{\multirow{2}{*}{MB/s}} &
     \multicolumn{1}{c|}{\multirow{2}{*}{GCt (s)}} &
     \multicolumn{2}{c|}{Minor} &
     \multicolumn{2}{c|}{Major} &
     \multicolumn{1}{c|}{\multirow{2}{*}{MB/s}} &
     \multicolumn{1}{c|}{\multirow{2}{*}{GCt (s)}}\\
\cline{3-4}\cline{7-10}\cline{13-16}
 & & \multicolumn{1}{c|}{\#} & \multicolumn{1}{c|}{\#} & & &
     \multicolumn{1}{c|}{\#} & \multicolumn{1}{c|}{\%} &
     \multicolumn{1}{c|}{\#} & \multicolumn{1}{c|}{\%} & & &
     \multicolumn{1}{c|}{\#} & \multicolumn{1}{c|}{\%} &
     \multicolumn{1}{c|}{\#} & \multicolumn{1}{c|}{\%} & & \\
\hline
\hline
cassandra$^\dagger$ & \multirow{12}{*}{Default} &
1 & 6 & 5.6 & 3.9 &
45 & 100 & 8 & 61 & 21.1 & 14.0 &
55 & 100 & 9 & 78 & 27.5 & 16.4\\
fop &  &
1 & 4 & 11.6 & 10.4 &
39 & 100 & 9 & 70 & 40.0 & 28.8 &
54 & 100 & 12 & 92 & 46.5 & 33.5\\
graphchi &  &
2 & 5 & 1.4 & 0.6 &
32 & 100 & 6 & 45 & 12.6 & 1.6 &
62 & 100 & 8 & 68 & 25.2 & 2.6\\
h2$^\dagger$ &  &
35 & 6 & 42.3 & 50.5 &
52 & 53 & 13 & 73 & 40.4 & 53.2 &
55 & 65 & 9 & 67 & 50.5 & 54.2\\
jython &  &
2 & 4 & 6.0 & 8.0 &
70 & 100 & 7 & 86 & 16.3 & 19.6 &
92 & 100 & 10 & 90 & 15.8 & 20.7\\
lusearch$^\dagger$ &  &
1 & 19 & 0.7 & 1.6 &
18 & 100 & 25 & 14 & 1.0 & 2.1 &
20 & 100 & 26 & 11 & 1.2 & 2.0\\
pmd &  &
7 & 5 & 24.8 & 4.7 &
42 & 100 & 8 & 60 & 90.6 & 15.9 &
56 & 100 & 10 & 58 & 110.2 & 22.2\\
spring$^\dagger$ &  &
11 & 10 & 1.6 & 3.7 &
128 & 100 & 13 & 49 & 3.1 & 8.3 &
942 & 100 & 18 & 83 & 3.5 & 27.9\\
sunflow &  &
13 & 20 & 3.9 & 2.0 &
29 & 88 & 24 & 14 & 5.7 & 2.4 &
33 & 98 & 26 & 16 & 5.3 & 2.4\\
tomcat$^\dagger$ &  &
0 & 10 & 0.8 & 2.9 &
217 & 100 & 12 & 92 & 1.3 & 7.5 &
587 & 100 & 24 & 97 & 1.6 & 15.2\\
xalan &  &
9 & 9 & 2.2 & 0.8 &
38 & 100 & 13 & 41 & 5.6 & 1.4 &
53 & 100 & 13 & 49 & 19.0 & 1.5\\
zxing &  &
2 & 4 & 65.8 & 1.7 &
33 & 100 & 10 & 56 & 94.9 & 2.9 &
43 & 100 & 11 & 55 & 98.9 & 3.2\\
\hline
cassandra$^\dagger$ & \multirow{10}{*}{Large} &
13 & 8 & 36.2 & 31.5 &
74 & 100 & 16 & 29 & 61.4 & 67.3 &
107 & 100 & 21 & 35 & 91.0 & 81.1\\
graphchi &  &
0 & 88 & 4.7 & 13.2 &
42 & 100 & 77 & 4 & 12.1 & 14.5 &
134 & 100 & 33 & 14 & 23.0 & 16.2\\
h2$^\dagger$ &  &
552 & 33 & 152.2 & 2922.7 &
573 & 5 & 38 & 18 & 149.8 & 2841.3 &
572 & 5 & 40 & 19 & 150.8 & 2850.4\\
jython &  &
38 & 7 & 23.6 & 105.7 &
213 & 100 & 28 & 38 & 24.7 & 160.0 &
622 & 100 & 74 & 8 & 25.6 & 264.2\\
lusearch$^\dagger$ &  &
0 & 20 & 0.2 & 40.9 &
4 & 100 & 108 & 1 & 0.2 & 41.6 &
10 & 100 & 210 & 1 & 0.2 & 42.0\\
pmd &  &
26 & 9 & 612.4 & 233.3 &
37 & 33 & 11 & 40 & 620.1 & 227.3 &
38 & 34 & 13 & 45 & 616.1 & 234.3\\
spring$^\dagger$ &  &
32 & 26 & 0.9 & 8.7 &
160 & 100 & 24 & 13 & 1.5 & 12.0 &
2818 & 100 & 30 & 89 & 1.7 & 65.8\\
sunflow &  &
11 & 51 & 3.9 & 3.9 &
26 & 80 & 58 & 9 & 4.6 & 4.5 &
36 & 98 & 60 & 7 & 4.5 & 4.7\\
tomcat$^\dagger$ &  &
0 & 31 & 0.5 & 6.3 &
326 & 100 & 10 & 92 & 0.5 & 10.1 &
1421 & 100 & 26 & 96 & 0.5 & 30.5\\
xalan &  &
11 & 18 & 1.9 & 1.4 &
45 & 100 & 14 & 14 & 3.2 & 1.5 &
49 & 100 & 14 & 16 & 3.7 & 1.5\\
\hline
SPECjbb2015 & Single JVM &
703 & 35 & 100 & 1684.2 & 744 & 25 & 37 & 15 & 105 & 1781.3 & 919 & 54 & 43 & 15 & 125 & 2314.9 \\
\hline
\end{tabular}
\caption{Baseline ZGC and OppZGC collection statistics. For each ZGC configuration, the columns show the number of minor and major collections, (for OppZGC only) the percentage of minor and major collections caused by the opportunistic rules, the data copy rate (i.e., MB copied for ZPage compaction or object promotions per second), and the aggregate CPU time of all ZGC workers in seconds. All configurations shown here use a maximum heap size of 32 GB.}
\label{tab:oppzgc_runtime}
\end{table*}

\subsubsection{DaCapo Chopin}
\label{sec:dacapo_exp}
DaCapo Chopin is a widely used Java benchmark suite consisting of real and open-source Java applications with a diverse range of computing behaviors, goals, and resource needs.
To focus our evaluation, we omit benchmarks that show no measurable steady-state performance difference with default input when the number of computing cores available to the application is reduced from 16 to 2.\footnote{Specifically, we use \texttt{numactl} to limit the CPU set available to the JVM process and set \texttt{-XX:ActiveProcessorCount} to the appropriate core count.}
We also omit tradebeans and tradesoap, which are not supported by our JDK (OpenJDK version 24)~\cite{tradebeans}.
Table~\ref{tab:dacapo} in Appendix~\ref{sec:dacapo_table} lists our selected DaCapo benchmarks along with performance and memory characteristics for the baseline ZGC configurations included in this study.

DaCapo includes multiple input sizes for most applications.
Our study uses the default input and large input (if one is included) for each selected benchmark, for a total of 22 inputs over 12 applications.
The DaCapo suite also includes a harness program that runs the benchmark for a given number of iterations or until convergence, which occurs when the standard deviation of the last three iterations divided by their average execution time is less than 3\%.
Since concurrent GC can potentially affect execution time of individual iterations (and thus convergence), we opted to set the number of iterations manually for each benchmark-input pair.
For the default inputs, we use a number of iterations that executes the workload for 30 seconds to 150 seconds in the baseline configuration.
All the large inputs were run with exactly 5 iterations per run.
\footnote{Backtesting against the DaCapo convergence formula shows that the number of iterations we used is sufficient for all but five benchmarks to converge with baseline ZGC: \emph{jython-default}, \emph{pmd-large}, \emph{sunflow-default}, \emph{sunflow-large}, and \emph{xalan-large}.}

For each benchmark, we report both \emph{startup time}, which is the execution time of the first iteration, and \emph{steady-state time}, which we compute as the average execution time of every iteration aside from the first two (i.e., the startup iteration plus one warmup iteration).
In this way, our steady-state measurements avoid the period with heaviest JIT compilation while still capturing GC costs over a common number of iterations.
Ten of the benchmark-input pairs, marked in Tables~\ref{tab:oppzgc_runtime} and \ref{tab:dacapo} with $^\dagger$, also report request-based latencies as their primary performance metric.
For these workloads, we report the simple (i.e., non-metered) p50, p90, p99, p99.9, and p99.99 latencies over the steady-state iterations.

\subsubsection{SPECjbb2015}
\label{sec:specjbb}
SPECjbb is a popular Java performance analysis benchmark for server platforms that scales with both compute and memory resources.
It includes options for running with multiple JVMs or across multiple hosts, but for this work, we run it using a single JVM in composite mode on our single node platform.

The workload models a supermarket IT infrastructure, including Point of Sale (POS) transactions, receipt and inventory management, interactions with suppliers, and data analytics.
The SPECjbb driver injects transactions at a specific rate called the Injection Rate (IR), and the system under test completes the transactions at a measured rate known as jOPS.
The full benchmark consists of five phases: High Bound Injection Rate (HBIR) Search, Response-Throughput (RT) curve building, Validation, Profiling, and Reporting.
Performance is determined by the first two phases.
The HBIR phase quickly estimates throughput capabilities of the system under test, and the RT curve building then walks the injection rate up more finely, locating the point where throughput stops scaling with injection rate, known as saturation.

The benchmark reports two primary performance measurements: the maximum jOPS, or throughput at the saturation point, and critical jOPS, which measures throughput with some acceptable latency under a range of service level agreements (SLAs).
Specifically, SPECjbb defines five SLAs at 10, 25, 50, 75, and 100 ms.
As the benchmark sweeps the injection rate, it finds the approximate throughput level at which the p99 response time crosses the SLA.
Critical jOPS is the geometric mean of those five throughput values.




\section{Evaluation}
\subsection{Selecting the Growth Ratio and Period}
As discussed in Section~\ref{sec:design}, OppZGC has two parameters that affect its scheduling decisions: the growth ratio $\alpha$ and the period $\delta$.
To select parameter values for this study, we conducted a sweep of different $\alpha$ values and $\delta$ values with the DaCapo benchmarks with their default inputs.
Specifically, we tested $\alpha$ values of 1.1, 1.0, 0.9, 0.8, 0.5, and 0.1 with $\delta$ values of $10\,\mathrm{ms}$, $20\,\mathrm{ms}$, $40\,\mathrm{ms}$, $80\,\mathrm{ms}$, and $160\,\mathrm{ms}$, for a total of 30 different configurations.
Of these, the three configurations that achieved the lowest maximum heap size without degrading the average steady-state performance were: ($\alpha$=1.0, $\delta=10\,\mathrm{ms}$), ($\alpha$=1.0, $\delta=20\,\mathrm{ms}$), and ($\alpha$=0.9, $\delta=20\,\mathrm{ms}$).
We then ran these three configurations with our full benchmark set.
This section shows detailed results for OppZGC with $\alpha$=1.0, $\delta$=20\,ms, which achieves the best average performance of the three configurations, and $\alpha$=0.9, $\delta$=20\,ms, which achieves the best average memory savings.

\subsection{OppZGC Collection Statistics}
Table~\ref{tab:oppzgc_runtime} presents collection statistics for baseline ZGC and two OppZGC configurations across our complete benchmark set.
As expected, both OppZGC configurations collect garbage more aggressively than baseline ZGC.
Most of these benchmarks leave at least some of the 16 available computing cores idle or underutilized during their execution.
To take advantage of the available CPU capacity, OppZGC runs from several dozen up to a few thousand additional collection cycles compared to baseline ZGC.

Increasing the number of collections also leads to higher data copy rates and requires additional CPU time to complete the extra collection work.
On average, OppZGC increases the CPU time spent on GC by $1.7\times$ with $\alpha=1.0$ and by $2.3\times$ with $\alpha=0.9$, compared to the baseline ZGC configuration.
However, the impact of OppZGC on GC effort can vary widely for different benchmarks.
For example, \emph{spring} exhibits GC CPU time increases ranging from $1.4\times$ to $7.6\times$ with OppZGC.
Other benchmarks, such as \emph{h2} and \emph{lusearch}, see little or no increase in GC activity with OppZGC.
For these benchmarks, the mutators exhaust CPU capacity throughout most of their execution, and thus, OppZGC forgoes additional collection cycles, thereby avoiding performance losses.
Notably, this extra collection work does not increase energy costs.
On average, package power is essentially unchanged ($<1\%$ increase), and package energy is reduced slightly (by $<2\%$, matching performance improvements) for the DaCapo benchmarks with OppZGC using either $\alpha=1.0$ or $\alpha=0.9$ and -Xmx32g.

\begin{figure*}[t]
  \centering 
  \begin{subfigure}{0.32\textwidth}\centering
    \includegraphics[width=\linewidth]{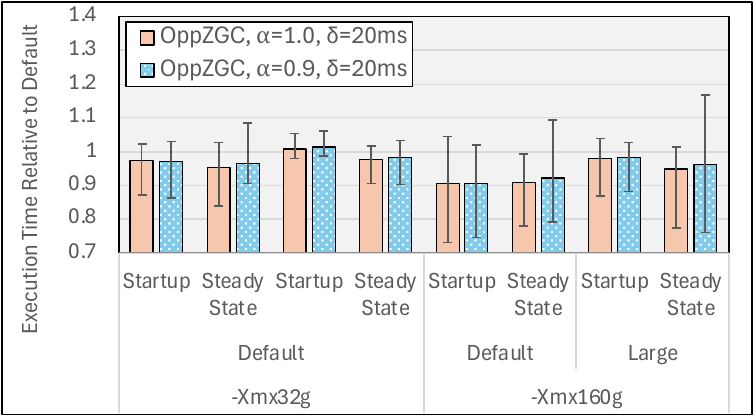}
    \caption{Startup and steady-state time.}\label{fig:dacapo_average_perf}
  \end{subfigure}\hfill
  \begin{subfigure}{0.32\textwidth}\centering
    \includegraphics[width=\linewidth]{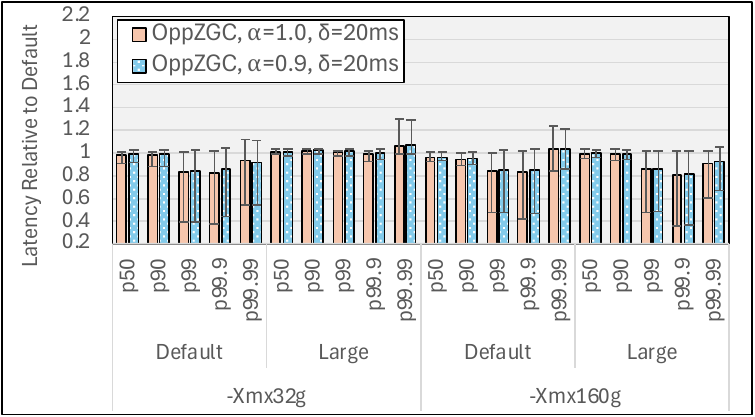} 
    \caption{Latency distribution.}\label{fig:dacapo_average_latency}
  \end{subfigure}\hfill
  \begin{subfigure}{0.32\textwidth}\centering
    \includegraphics[width=\linewidth]{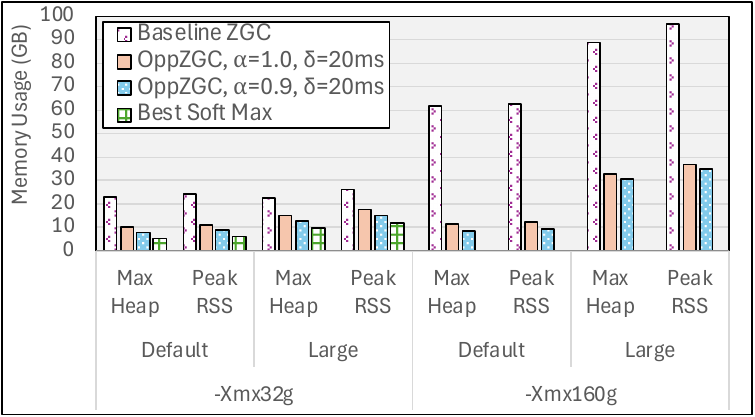}
    \caption{Maximum heap used and peak RSS.}\label{fig:dacapo_average_memory}
  \end{subfigure}
  \caption{OppZGC performance and memory usage, averaged over the DaCapo benchmark groups. (a) and (b) show results relative to baseline ZGC and include error bars showing the best and worst result from each group (lower is better). Results in~(c) are in GB. The -Xmx32g group also shows results for the best per-benchmark soft max configuration.}
  \Description{Startup and steady-state time, latency distribution, and memory usage for OppZGC averaged over the DaCapo benchmark groups.}
  \label{fig:dacapo_avg}
\end{figure*}
\subsection{Evaluation with 16 Cores}
\label{sec:eval_16-core}
\subsubsection{DaCapo Chopin Performance}
Figures~\ref{fig:dacapo_average_perf} and \ref{fig:dacapo_average_latency} present the average performance of OppZGC in terms of startup and steady-state execution time, as well as the p50, p90, p99, p99.9, and p99.99 latencies.
Both panels show results for the two OppZGC configurations relative to baseline ZGC performance for the same input and -Xmx value.
Error bars show the best (low end) and worst (high end) performing benchmark for each configuration.
Detailed per-benchmark results are also presented in Appendix~\ref{sec:eval_varying_cores}.

Our results show that both OppZGC configurations achieve similar startup and steady-state performance to baseline ZGC in most cases, with only a few negative exceptions.
Even when the CPU is underutilized, increasing the rate of collection may still slow mutator progress because doing so increases the frequency of barrier slow paths and can evict warm data from processor caches.
However, we find that mutator interference is somewhat limited for both OppZGC configurations.
In some cases, startup time appears to degrade by 10\% or more, but run-to-run variance for startup time is relatively high and these differences do not fall outside the 95\% CIs.
In the worst case, steady-state execution time increases by about 6\% for \emph{xalan-large} with -Xmx32g.

On the other hand, several benchmarks exhibit more substantial startup and steady-state performance improvements compared to baseline ZGC.
In the best cases, individual benchmarks can improve by more than 20\% in startup (\emph{xalan-default}) or steady-state (\emph{pmd-default}) performance.
Similar to the soft max results in Section~\ref{sec:hardmax_softmax}, we find that these performance improvements are primarily driven by reduced paging overhead: fewer page faults enable better startup performance, and a smaller heap footprint improves TLB efficiency.
For the default inputs with a 160 GB heap limit, these effects improve the startup and steady-state performance by about 9\%, on average.

Request latencies from p50 up through p99.9 are essentially unharmed by either OppZGC configuration.
However, we find that OppZGC can degrade the p99.99 tail latency, in a few cases.
With -Xmx32g, both OppZGC configurations increase p99.99 latency for \emph{cassandra-large} by about 30\%, and with the larger heap maximum, OppZGC increases p99.99 latencies for \emph{h2-default} by between 16\% and 24\% and for \emph{lusearch-default} by 15\% to 21\%. 
Despite these few exceptions, OppZGC achieves typical and tail latencies very similar to baseline ZGC, and for the vast majority of workloads and execution scenarios that we tested, it preserves the low-latency behavior that ZGC aims to provide.

\subsubsection{DaCapo Chopin Memory}
\label{sec:dacapo_16_eval}
Figure~\ref{fig:dacapo_average_memory} presents the average memory usage of baseline ZGC and the two OppZGC configurations in terms of maximum heap used and peak RSS, both in GB.
Results are grouped by benchmark input and maximum heap size.
For comparison, we plot an additional bar in the -Xmx32g results based on our experiments varying the soft max parameter in Section~\ref{sec:hardmax_softmax}.
Specifically, this bar shows the average memory usage of the per-benchmark soft max value that exhibits the lowest heap usage without hurting any performance metric (i.e., startup, steady-state, or latency) relative to baseline ZGC.

OppZGC substantially reduces memory usage for at least 10 (of 12) default input benchmarks (depending on $\alpha$ and the hard heap limit) and for at least 6 (of 10) of the large input benchmarks compared to baseline ZGC.
With -Xmx32g, OppZGC reduces maximum heap usage by 61\%, on average, with $\alpha=1.0$ and by 75\% with $\alpha=0.9$; with -Xmx160g, the reductions are 86\% and 90\%, respectively.\footnote{These percentages are computed for each benchmark as the reduction relative to baseline ZGC, then averaged over the full group. Since Figure~\ref{fig:dacapo_average_memory} presents the mean heap occupancy of each configuration as a scalar, its implied ratios show a similar trend, but do not match these percentages exactly.}
As expected, the benchmarks where OppZGC is least effective at reducing memory usage are the same benchmarks with highest CPU utilization in the baseline configuration.
For these benchmarks, OppZGC exhibits similar memory usage to baseline ZGC because it still checks the same scheduling rules as baseline ZGC, even if the opportunistic rules never fire.


The results allow two other notable observations.
First, reducing heap usage produces similar savings in resident set size.
Although other factors, such as stack memory and the distribution of objects on ZPages, can increase peak RSS for some workloads, the impact of these factors is relatively small for these benchmarks.
Additionally, while OppZGC substantially reduces memory usage compared to baseline ZGC, a well-tuned soft maximum can still achieve better memory savings for 13 of our 22 benchmarks.
On average, the best soft max configuration uses 2.8 GB less heap memory than OppZGC with $\alpha=0.9$ (with -Xmx32g).
However, these additional memory savings come at the cost of offline, per-benchmark tuning, which is impractical for many applications.
OppZGC achieves these improvements using the same configuration for the entire benchmark set, and does not require customization or tuning for each workload.

\begin{figure}
    \centering
    
    \begin{subfigure}[b]{0.48\columnwidth}
        \includegraphics[width=\linewidth, page=2]{figures/specjbb.pdf}
        \caption{Max jOPS.}
        \label{fig:specjbb-max-jops}
    \end{subfigure}
    \hfill
    \begin{subfigure}[b]{0.48\columnwidth}
        \includegraphics[width=\linewidth, page=1]{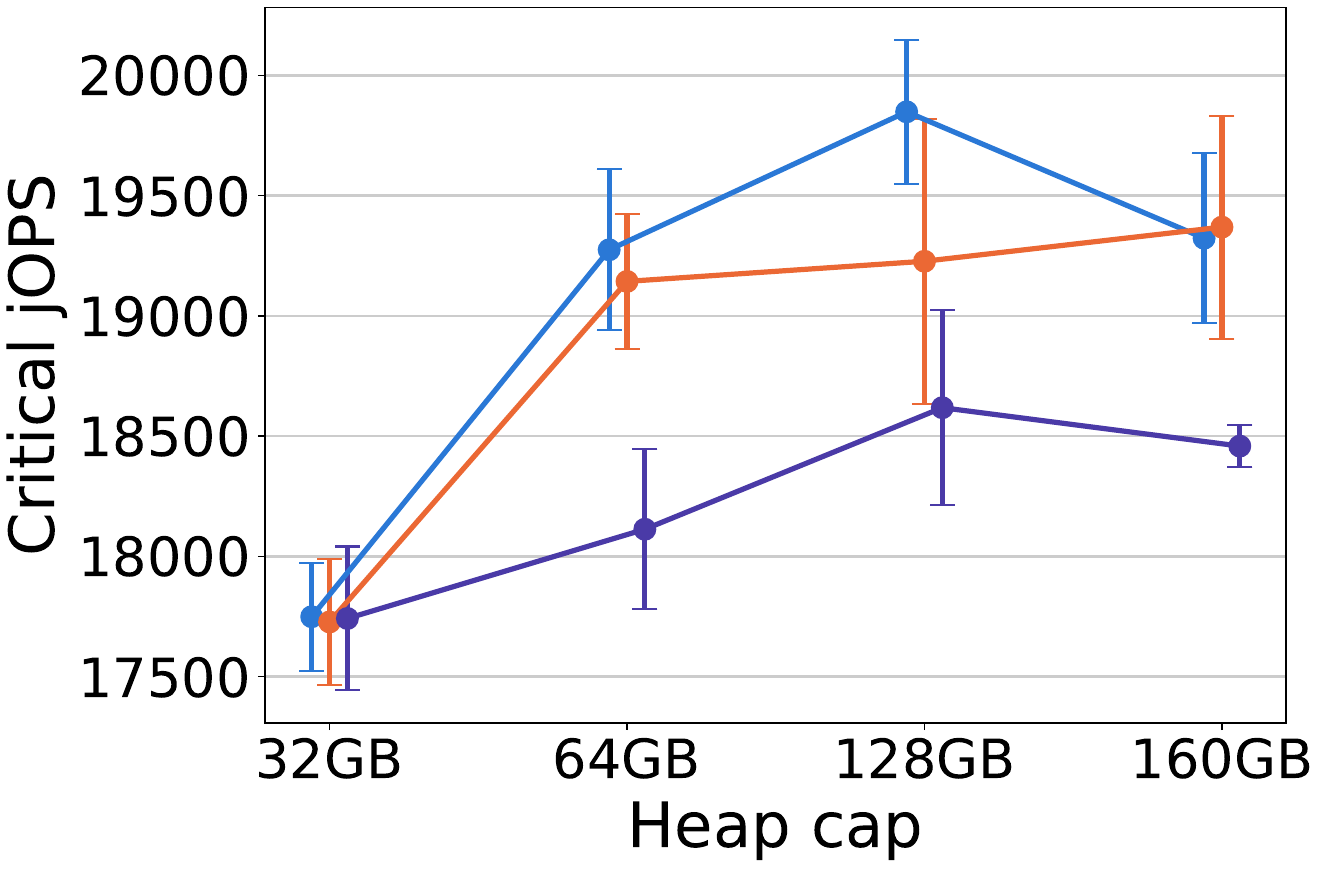}
        \caption{Critical jOPS.}
        \label{fig:specjbb-critical-jops}
    \end{subfigure}

    \begin{subfigure}[b]{0.48\columnwidth}
        \includegraphics[width=\linewidth, page=3]{figures/specjbb.pdf}
        \caption{Max heap used (GB).}
        \label{fig:specjbb-max-heap}
    \end{subfigure}
    \hfill
    \begin{subfigure}[b]{0.48\columnwidth}
        \includegraphics[width=\linewidth, page=4]{figures/specjbb.pdf}
        \caption{Peak RSS (GB).}
        \label{fig:specjbb-max-rss}
    \end{subfigure}

    \caption{SPECjbb performance and memory usage with OppZGC and baseline ZGC. Each line shows results with different -Xmx limits: 32g, 64g, 128g, and 160g.}
    \Description{Line graphs showing memory and performance characteristics of SPECjbb with Baseline ZGC and OppZGC with different maximum heap sizes.}
    \label{fig:specjbb_eval}
\end{figure}
\begin{figure*}[t]
  \centering
  \begin{subfigure}{0.32\textwidth}\centering
    \includegraphics[width=\linewidth]{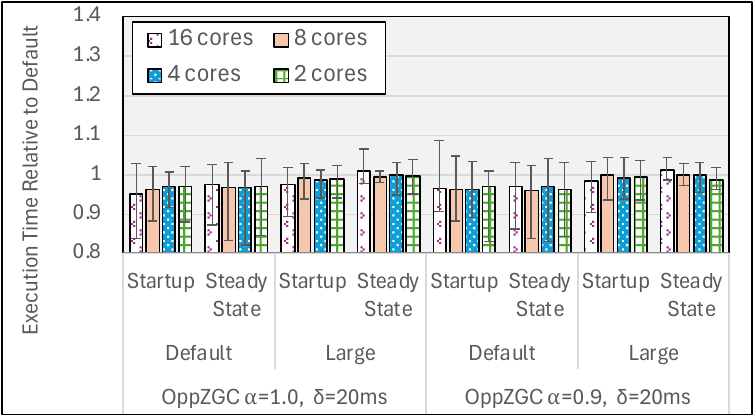}
    \caption{Startup and steady-state time.}\label{fig:oppzgc_core_perf}
  \end{subfigure}\hfill
  \begin{subfigure}{0.32\textwidth}\centering
    \includegraphics[width=\linewidth]{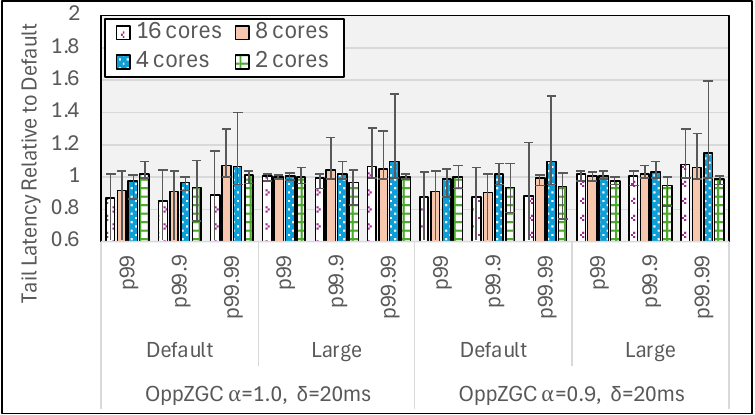}
    \caption{Tail latencies.}\label{fig:oppzgc_core_latency}
  \end{subfigure}\hfill
  \begin{subfigure}{0.32\textwidth}\centering
    \includegraphics[width=\linewidth]{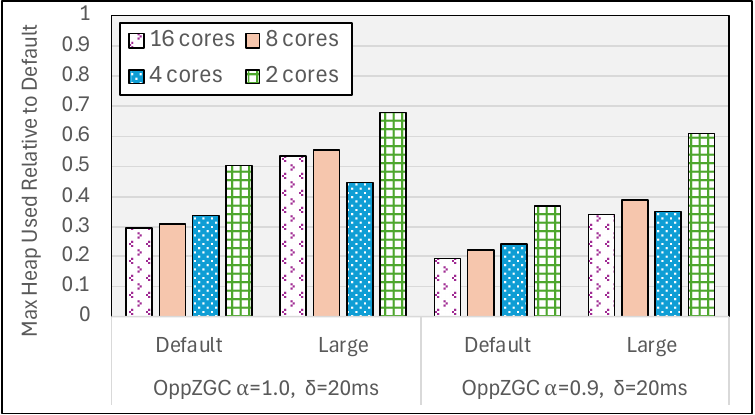}
    \caption{Maximum heap occupancy.}\label{fig:oppzgc_core_memory}
  \end{subfigure}
  \caption{OppZGC with varying CPU capacity, relative to baseline ZGC at the same capacity, averaged over the DaCapo benchmark groups. Error bars in~(a) and (b) show the best and worst result from each group (lower is better).}
  \Description{Startup and steady-state time, tail latencies, and memory usage for OppZGC with varying CPU capacity.}
  \label{fig:oppzgc_core}
\end{figure*}
\subsubsection{SPECjbb2015}
Since SPECjbb performance typically scales with the availability of memory resources, we evaluated OppZGC with SPECjbb with four different maximum heap sizes: 32 GB, 64 GB, 128 GB, and 160 GB.
Figure~\ref{fig:specjbb_eval} presents the max jOPS (maximum throughput), critical jOPS (throughput achieved given some latency constraints), maximum heap usage, and peak RSS for the baseline and two OppZGC configurations (with $\alpha=1.0$ and $\alpha=0.9$) with each maximum heap size.

As with DaCapo, the results show that both OppZGC configurations have little to no impact on peak throughput, regardless of the maximum heap size.
However, the more aggressive OppZGC configuration can increase request latency in some cases, reducing the critical jOPS.
In the worst case, OppZGC with $\alpha=0.9$ and a 128 GB maximum heap size reduces critical jOPS by 6.2\% compared to baseline ZGC.
The more conservative $\alpha=1.0$ setting eliminates this degradation and achieves essentially the same latency-constrained throughput as baseline ZGC.

While OppZGC has minimal impact on SPECjbb performance, it still reduces memory usage substantially compared to baseline ZGC.
In this case, there is also very little difference in memory usage between the two OppZGC configurations, with $\alpha=0.9$ using slightly less memory at most maximum heap sizes.
For $\alpha=1.0$, reductions in maximum heap occupancy range from 19.4\% with -Xmx64g to 21.3\% with -Xmx128g, which corresponds to reductions in peak RSS ranging from 15\% to 20\%.
While these reductions are not as pronounced as the average savings with DaCapo, they demonstrate that OppZGC can even reduce memory usage for CPU-intensive workloads with efficient scaling.

\vspace{-1ex}
\subsection{Evaluation with Varying CPU Capacity}

\subsubsection{Performance with Varying CPU Capacity}
To evaluate OppZGC with varying CPU capacity, we conducted a series of experiments with the Java process restricted to use only 8, 4, and 2 of the computing cores on our Intel-based platform.
Figures~\ref{fig:oppzgc_core_perf} and \ref{fig:oppzgc_core_latency} show the average performance in terms of startup time, steady-state time, and tail latencies for each core restriction.
The panels show average results, grouped by input size, for the two OppZGC configurations with different core restrictions relative to a baseline ZGC configuration \emph{with the same core restriction}.
All results use a maximum heap size of 32 GB.
Similar to Figures~\ref{fig:dacapo_average_perf} and \ref{fig:dacapo_average_latency}, error bars show the best (low end) and worst (high end) performing benchmark for each configuration.
For added context, each panel also reprints the results for the 16-core configuration from Section~\ref{sec:eval_16-core}.
Additionally, since typical latencies (i.e., p50 and p90) are essentially unaffected by OppZGC, Figure~\ref{fig:oppzgc_core_latency} omits these results and plots only the p99, p99.9, and p99.99 tail latencies to reduce clutter in the graph. 
Full results are available in Appendix~\ref{sec:eval_varying_cores}.

In general, reducing the number of computing cores available does not change the relative startup or steady-state performance of OppZGC.
On average, each OppZGC configuration exhibits startup and steady-state performance that is about the same or better than baseline ZGC with the same number of cores.
In contrast, there are a few cases where reducing the available CPU capacity causes additional degradations to the tail latency, especially at p99.99.
In the worst cases, OppZGC increases p99.99 latency for \emph{cassandra} by 40\% to 57\%, depending on the benchmark input and $\alpha$ setting, with execution restricted to only four computing cores.

Notably, the most pronounced degradations all occur when the application is limited to 4 computing cores, rather than 8 or 2.
With only two cores available, both cores are almost always saturated with mutator activity for many of the benchmarks (which prevents the opportunistic rules from firing), while with 16 and 8 cores available, at least some CPU is left unused for most of the run for most benchmarks, which allows for additional collection with minimal cost.
However, with only four cores available, several of our benchmarks, and especially \emph{cassandra}, oscillate between periods where the CPUs are underutilized and then quickly become entirely saturated.
Thus, there are more cases where the OppZGC rules incorrectly predict that there is sufficient CPU capacity to support a collection cycle.
For these cases, a more conservative OppZGC configuration with a longer period ($\delta$) can help mitigate these performance losses.
For \emph{cassandra} with four cores, we found that lengthening the period to a much more conservative 10\,s eliminates most tail latency degradations (though the large input still exhibits a 25\% increase in p99.99 latency), while still reducing maximum heap usage by 57\% and 76\% for the default and large inputs.

\subsubsection{Memory Usage with Varying CPU Capacity}
Finally, Figure~\ref{fig:oppzgc_core_memory} presents the maximum heap occupancy for the two OppZGC configurations relative to baseline ZGC for each core configuration, averaged over each benchmark-input set.
We find that limiting the workload to only 8 or 4 cores has only a minor impact on the average memory savings.
In these configurations, OppZGC still frequently predicts (sometimes incorrectly, as discussed above) there is enough CPU capacity to support additional collection cycles for most workloads.
However, with only two cores for the entire process, OppZGC becomes more conservative and more often forgoes collections to allow the mutators to operate unimpeded.
Overall, when execution is restricted to only two cores, OppZGC with $\alpha=0.9$ reduces the maximum heap occupancy for 7 of 12 default input benchmarks and for 3 of 10 large input benchmarks, resulting in average reductions in heap occupancy of 63\% and 39\% for the two input sets, respectively.
In contrast, the same OppZGC configuration with 16-core execution reduces heap memory usage by 80\% and 65\% for each input set.

\section{Related Work}
Since the introduction of tracing garbage collectors, which defer reclamation of dead objects rather than freeing them immediately~\cite{mccarthyRecursiveFunctionsSymbolic1960}, researchers have sought to understand how to balance memory usage with collection costs.
Several works have studied and quantified the computational costs of different types of collectors, including generational collectors~\cite{hertzQuantifyingPerformanceGarbage,blackburn_myths_2004,appel1989spe,lieberman1983cacm}.
Bacon et al. proposed a unified cost model for trace-based and reference-counting collectors and studied tradeoffs across different designs~\cite{bacon2004oopsla}.
Wilson~\cite{wilsonUniprocessorGarbageCollection} surveyed the canonical collection algorithms and described their tradeoffs, including Baker's algorithm~\cite{baker1978}, which is the basis of ZGC's load barriers.
Yang and Wrigstad also provided a comprehensive description of an earlier version of ZGC (from JDK 15)~\cite{yang_deep_2022}.

Several prior works have observed that effective collection can reduce paging costs and have designed collection strategies to exploit this consequence~\cite{alonsoAdvisorFlexibleWorking1990,yangAutomaticHeapSizing2004,yang2006osdi,grzegorczykIslaVistaHeap,brunoDynamicVerticalMemory}.
Kolokasis et al.~\cite{kolokasisFlexHeapDynamicOAware2026} balance GC and I/O cache allocation, partitioning a fixed DRAM budget between the heap and the page cache using the CPU time lost to GC and to I/O stalls.
White et al.~\cite{whiteControlTheoryPrincipled2013} employ a principled mathematical approach, tuning a PID controller to track a heap size more responsively than the heuristic-based mechanisms of Jikes and HotSpot.
In contrast to OppZGC, each of these works sets the heap size directly as a means to control GC cost.
Moreover, these works target collectors that serialize reclamation with the mutators, where the costs are paid in pause frequency and duration.
Since concurrent collectors do most of their work as the mutators run, contention for the CPU can worsen mutator performance.
OppZGC aims to avoid this contention by only running extra collections in otherwise idle CPU capacity.

Perhaps most closely related to this study are prior works that use system load to influence collector scheduling.
Zhao et al.~\cite{zhaoImprovingConcurrentGC2022} schedule concurrent work into troughs in system load to protect latency-critical tenants co-located with best-effort jobs.
Shimchenko et al.~\cite{shimchenko_monk_2025} schedule in troughs while also building priority-inversion-aware slack scheduling directly into the JVM.
Both of these works use the same signal of system load as OppZGC, but they use this information for scheduling collections to reduce latency instead of increasing collection frequency to restrict memory usage.

Another important related effort, which has since been adopted by a JEP for a future OpenJDK release~\cite{osterlund_jep_2026}, schedules collections by targeting a ratio of collector to mutator CPU time and letting heap size follow~\cite{tavakolisomeh_heap_2023}.
This ratio, however, is independent of CPU capacity.
It aims for the same target whether the CPU is saturated or idle, even though the marginal cost of collection is contention for CPU.
The policy is thus vulnerable to scenarios where the available CPU capacity shifts during execution and becomes misaligned with the chosen target ratio.
In contrast, OppZGC triggers extra collections only when sufficient idle CPU capacity is detected.


\section{Conclusion}
This work presents Opportunistic ZGC: a feedback-directed scheduling policy for concurrent collectors that exploits otherwise unused CPU capacity to constrain the heap, while keeping the costs of extra collection cycles low.
While the approach includes two tunable parameters, this study finds parameter values that reduce memory usage for a diverse set of benchmarks.
Thus, this work fills a gap between current practices of relatively conservative GC scheduling, which permits the heap to expand to use the space allowed by a predefined maximum, and per-application tuning, which allows users to constrain the heap with knobs for setting a soft heap maximum or a target ratio for collector-to-mutator CPU time.
On our 16-core Intel platform, OppZGC reduces maximum heap usage for our DaCapo benchmarks by between 61\% and 90\%, on average, compared to default ZGC, depending on configuration.
For SPECjbb, OppZGC reduces memory usage by up to 21\% in large memory configurations, with no cost to maximum or latency-constrained throughput.

In scenarios with very limited CPU capacity, OppZGC automatically adjusts GC scheduling to avoid stealing scarce CPU from mutator threads.
In some configurations, bursty workloads can cause OppZGC to mispredict the CPU capacity available for collection, which can hurt request latency performance in the extreme tail.
Future efforts will focus on extending OppZGC to identify scenarios where opportunistic collection cycles can hurt performance and designing strategies to forgo these extra collections or mitigate their costs.
However, for the vast majority of applications and execution scenarios, this work shows that OppZGC effectively constrains the application heap, with minimal harm to startup time, throughput, and request latencies.

\begin{acks}
This research was supported by the National Science Foundation under awards CNS-1943305 and 2514057. AI coding tools assisted with software development tasks, including patches to HotSpot and scripts for running experiments and summarizing results. Large language model tools were used in a limited way to suggest wording revisions. The authors wrote the initial draft, verified the cited literature, and made all final decisions about wording and content. No AI tools were used to generate the scientific ideas, experimental methodology, organization of results, or conclusions.
\end{acks}
\clearpage

\bibliographystyle{ACM-Reference-Format}
\bibliography{Opportunistic_GC}


\begin{thebibliography}{31}


\ifx \showCODEN    \undefined \def \showCODEN     #1{\unskip}     \fi
\ifx \showISBNx    \undefined \def \showISBNx     #1{\unskip}     \fi
\ifx \showISBNxiii \undefined \def \showISBNxiii  #1{\unskip}     \fi
\ifx \showISSN     \undefined \def \showISSN      #1{\unskip}     \fi
\ifx \showLCCN     \undefined \def \showLCCN      #1{\unskip}     \fi
\ifx \shownote     \undefined \def \shownote      #1{#1}          \fi
\ifx \showarticletitle \undefined \def \showarticletitle #1{#1}   \fi
\ifx \showURL      \undefined \def \showURL       {\relax}        \fi
\providecommand\bibfield[2]{#2}
\providecommand\bibinfo[2]{#2}
\providecommand\natexlab[1]{#1}
\providecommand\showeprint[2][]{arXiv:#2}

\bibitem[Alonso and Appel(1990)]%
        {alonsoAdvisorFlexibleWorking1990}
\bibfield{author}{\bibinfo{person}{Rafael Alonso} {and}
  \bibinfo{person}{Andrew~W. Appel}.} \bibinfo{year}{1990}\natexlab{}.
\newblock \showarticletitle{An {Advisor} for {Flexible} {Working} {Sets}}. In
  \bibinfo{booktitle}{\emph{Proceedings of the 1990 {ACM} {SIGMETRICS}
  {Conference} on {Measurement} and {Modeling} of {Computer} {Systems}}}
  \emph{(\bibinfo{series}{{SIGMETRICS} '90})}. \bibinfo{publisher}{Association
  for Computing Machinery}, \bibinfo{address}{New York, NY, USA},
  \bibinfo{pages}{153--162}.
\newblock
\href{https://doi.org/10.1145/98457.98753}{doi:\nolinkurl{10.1145/98457.98753}}


\bibitem[Appel(1989)]%
        {appel1989spe}
\bibfield{author}{\bibinfo{person}{Andrew~W. Appel}.}
  \bibinfo{year}{1989}\natexlab{}.
\newblock \showarticletitle{Simple Generational Garbage Collection and Fast
  Allocation}.
\newblock \bibinfo{journal}{\emph{Software: Practice and Experience}}
  \bibinfo{volume}{19}, \bibinfo{number}{2} (\bibinfo{year}{1989}),
  \bibinfo{pages}{171--183}.
\newblock
\href{https://doi.org/10.1002/spe.4380190206}{doi:\nolinkurl{10.1002/spe.4380190206}}


\bibitem[Bacon et~al\mbox{.}(2004)]%
        {bacon2004oopsla}
\bibfield{author}{\bibinfo{person}{David~F. Bacon}, \bibinfo{person}{Perry
  Cheng}, {and} \bibinfo{person}{V.~T. Rajan}.}
  \bibinfo{year}{2004}\natexlab{}.
\newblock \showarticletitle{A Unified Theory of Garbage Collection}. In
  \bibinfo{booktitle}{\emph{Proceedings of the 19th Annual ACM SIGPLAN
  Conference on Object-Oriented Programming, Systems, Languages, and
  Applications (OOPSLA)}}. \bibinfo{publisher}{ACM}, \bibinfo{pages}{50--68}.
\newblock
\href{https://doi.org/10.1145/1028976.1028982}{doi:\nolinkurl{10.1145/1028976.1028982}}


\bibitem[Baker(1978)]%
        {baker1978}
\bibfield{author}{\bibinfo{person}{Henry~G. Baker}.}
  \bibinfo{year}{1978}\natexlab{}.
\newblock \showarticletitle{List Processing in Real Time on a Serial Computer}.
\newblock \bibinfo{journal}{\emph{Commun. ACM}} \bibinfo{volume}{21},
  \bibinfo{number}{4} (\bibinfo{date}{April} \bibinfo{year}{1978}),
  \bibinfo{pages}{280--294}.
\newblock
\href{https://doi.org/10.1145/359460.359470}{doi:\nolinkurl{10.1145/359460.359470}}


\bibitem[Blackburn et~al\mbox{.}(2025)]%
        {dacapo2025asplos}
\bibfield{author}{\bibinfo{person}{Stephen~M. Blackburn},
  \bibinfo{person}{Zixian Cai}, \bibinfo{person}{Rui Chen}, \bibinfo{person}{Xi
  Yang}, \bibinfo{person}{John Zhang}, {and} \bibinfo{person}{John~N. Zigman}.}
  \bibinfo{year}{2025}\natexlab{}.
\newblock \showarticletitle{Rethinking {Java} {Performance} {Analysis}}. In
  \bibinfo{booktitle}{\emph{Proceedings of the 30th {ACM} {International}
  {Conference} on {Architectural} {Support} for {Programming} {Languages} and
  {Operating} {Systems}}} \emph{(\bibinfo{series}{{ASPLOS} '25})}.
  \bibinfo{publisher}{Association for Computing Machinery},
  \bibinfo{address}{New York, NY, USA}, \bibinfo{pages}{940--954}.
\newblock
\href{https://doi.org/10.1145/3669940.3707217}{doi:\nolinkurl{10.1145/3669940.3707217}}


\bibitem[Blackburn et~al\mbox{.}(2004)]%
        {blackburn_myths_2004}
\bibfield{author}{\bibinfo{person}{Stephen~M. Blackburn},
  \bibinfo{person}{Perry Cheng}, {and} \bibinfo{person}{Kathryn~S. McKinley}.}
  \bibinfo{year}{2004}\natexlab{}.
\newblock \showarticletitle{Myths and {Realities}: {The} {Performance} {Impact}
  of {Garbage} {Collection}}. In \bibinfo{booktitle}{\emph{{ACM} {SIGMETRICS}
  {International} {Conference} on {Measurement} and {Modeling} of {Computer}
  {Systems}}} \emph{(\bibinfo{series}{{ACM} {SIGMETRICS} {Performance}
  {Evaluation} {Review} 32(1)})}. \bibinfo{publisher}{ACM Press},
  \bibinfo{pages}{25--36}.
\newblock
\href{https://doi.org/10.1145/1005686.1005693}{doi:\nolinkurl{10.1145/1005686.1005693}}


\bibitem[Bruno et~al\mbox{.}(2018)]%
        {brunoDynamicVerticalMemory}
\bibfield{author}{\bibinfo{person}{Rodrigo Bruno}, \bibinfo{person}{Paulo
  Ferreira}, \bibinfo{person}{Ruslan Synytsky}, \bibinfo{person}{Tetiana
  Fydorenchyk}, \bibinfo{person}{Jia Rao}, \bibinfo{person}{Hang Huang}, {and}
  \bibinfo{person}{Song Wu}.} \bibinfo{year}{2018}\natexlab{}.
\newblock \showarticletitle{Dynamic {Vertical} {Memory} {Scalability} for
  {OpenJDK} {Cloud} {Applications}}. In \bibinfo{booktitle}{\emph{Proceedings
  of the 2018 {ACM} {SIGPLAN} {International} {Symposium} on {Memory}
  {Management}}} \emph{(\bibinfo{series}{{ISMM} 2018})}.
  \bibinfo{publisher}{Association for Computing Machinery},
  \bibinfo{address}{New York, NY, USA}, \bibinfo{pages}{59--70}.
\newblock
\href{https://doi.org/10.1145/3210563.3210567}{doi:\nolinkurl{10.1145/3210563.3210567}}


\bibitem[{DaCapo Benchmark Suite}(2025)]%
        {tradebeans}
\bibfield{author}{\bibinfo{person}{{DaCapo Benchmark Suite}}.}
  \bibinfo{year}{2025}\natexlab{}.
\newblock \bibinfo{title}{tradebeans and tradesoap fail on {JDK} 24}.
\newblock
\shownote{DaCapo benchmark suite issue \#350; WildFly 26, used by the Trade*
  workloads, is incompatible with Java releases newer than 21}.
\newblock
\urldef\tempurl%
\url{https://github.com/dacapobench/dacapobench/issues/350}
\showURL{%
\tempurl}


\bibitem[Georges et~al\mbox{.}(2007)]%
        {georges2007oopsla}
\bibfield{author}{\bibinfo{person}{Andy Georges}, \bibinfo{person}{Dries
  Buytaert}, {and} \bibinfo{person}{Lieven Eeckhout}.}
  \bibinfo{year}{2007}\natexlab{}.
\newblock \showarticletitle{Statistically {Rigorous} {Java} {Performance}
  {Evaluation}}. In \bibinfo{booktitle}{\emph{Proceedings of the 22nd {ACM}
  {SIGPLAN} {Conference} on {Object}-{Oriented} {Programming} {Systems},
  {Languages} and {Applications}}} \emph{(\bibinfo{series}{{OOPSLA} '07})}.
  \bibinfo{publisher}{Association for Computing Machinery},
  \bibinfo{address}{New York, NY, USA}, \bibinfo{pages}{57--76}.
\newblock
\href{https://doi.org/10.1145/1297027.1297033}{doi:\nolinkurl{10.1145/1297027.1297033}}


\bibitem[Grzegorczyk et~al\mbox{.}(2007)]%
        {grzegorczykIslaVistaHeap}
\bibfield{author}{\bibinfo{person}{Chris Grzegorczyk}, \bibinfo{person}{Sunil
  Soman}, \bibinfo{person}{Chandra Krintz}, {and} \bibinfo{person}{Rich
  Wolski}.} \bibinfo{year}{2007}\natexlab{}.
\newblock \showarticletitle{Isla {Vista} {Heap} {Sizing}: {Using} {Feedback} to
  {Avoid} {Paging}}. In \bibinfo{booktitle}{\emph{Proceedings of the
  {International} {Symposium} on {Code} {Generation} and {Optimization} ({CGO}
  '07)}}. \bibinfo{publisher}{IEEE Computer Society},
  \bibinfo{pages}{325--340}.
\newblock
\href{https://doi.org/10.1109/CGO.2007.20}{doi:\nolinkurl{10.1109/CGO.2007.20}}


\bibitem[Hertz and Berger(2005)]%
        {hertzQuantifyingPerformanceGarbage}
\bibfield{author}{\bibinfo{person}{Matthew Hertz} {and}
  \bibinfo{person}{Emery~D. Berger}.} \bibinfo{year}{2005}\natexlab{}.
\newblock \showarticletitle{Quantifying the {Performance} of {Garbage}
  {Collection} vs. {Explicit} {Memory} {Management}}. In
  \bibinfo{booktitle}{\emph{Proceedings of the 20th {ACM} {SIGPLAN}
  {Conference} on {Object}-{Oriented} {Programming}, {Systems}, {Languages},
  and {Applications}}} \emph{(\bibinfo{series}{{OOPSLA} '05})}.
  \bibinfo{publisher}{Association for Computing Machinery},
  \bibinfo{address}{New York, NY, USA}, \bibinfo{pages}{313--326}.
\newblock
\href{https://doi.org/10.1145/1094811.1094836}{doi:\nolinkurl{10.1145/1094811.1094836}}


\bibitem[Kolokasis et~al\mbox{.}(2026)]%
        {kolokasisFlexHeapDynamicOAware2026}
\bibfield{author}{\bibinfo{person}{Iacovos~G. Kolokasis},
  \bibinfo{person}{Shoaib Akram}, \bibinfo{person}{Foivos~S. Zakkak},
  \bibinfo{person}{Polyvios Pratikakis}, {and} \bibinfo{person}{Angelos
  Bilas}.} \bibinfo{year}{2026}\natexlab{}.
\newblock \showarticletitle{{FlexHeap}: {Dynamic} {I}/{O}-{Aware} {Heap}
  {Resizing} for {Managed} {Applications}}.
\newblock \bibinfo{journal}{\emph{Proceedings of the ACM on Programming
  Languages}} \bibinfo{volume}{10}, \bibinfo{number}{PLDI}
  (\bibinfo{date}{June} \bibinfo{year}{2026}), \bibinfo{pages}{5--28}.
\newblock
\href{https://doi.org/10.1145/3808247}{doi:\nolinkurl{10.1145/3808247}}


\bibitem[Lidén and Karlsson(2018)]%
        {liden_jep_2018}
\bibfield{author}{\bibinfo{person}{Per Lidén} {and} \bibinfo{person}{Stefan
  Karlsson}.} \bibinfo{year}{2018}\natexlab{}.
\newblock \bibinfo{booktitle}{\emph{{JEP} 333: {ZGC}: {A} {Scalable}
  {Low}-{Latency} {Garbage} {Collector} ({Experimental})}}.
\newblock \bibinfo{type}{{T}echnical {R}eport}. \bibinfo{institution}{OpenJDK}.
\newblock
\urldef\tempurl%
\url{http://openjdk.java.net/jeps/333}
\showURL{%
\tempurl}


\bibitem[Lieberman and Hewitt(1983)]%
        {lieberman1983cacm}
\bibfield{author}{\bibinfo{person}{Henry Lieberman} {and} \bibinfo{person}{Carl
  Hewitt}.} \bibinfo{year}{1983}\natexlab{}.
\newblock \showarticletitle{A Real-Time Garbage Collector Based on the
  Lifetimes of Objects}.
\newblock \bibinfo{journal}{\emph{Commun. ACM}} \bibinfo{volume}{26},
  \bibinfo{number}{6} (\bibinfo{date}{June} \bibinfo{year}{1983}),
  \bibinfo{pages}{419--429}.
\newblock
\href{https://doi.org/10.1145/358141.358147}{doi:\nolinkurl{10.1145/358141.358147}}


\bibitem[{Linux man-pages project}(2024)]%
        {getrusage}
\bibfield{author}{\bibinfo{person}{{Linux man-pages project}}.}
  \bibinfo{year}{2024}\natexlab{}.
\newblock \bibinfo{title}{getrusage(2) --- {Linux} manual page}.
\newblock
\urldef\tempurl%
\url{https://man7.org/linux/man-pages/man2/getrusage.2.html}
\showURL{%
\tempurl}


\bibitem[McCarthy(1960)]%
        {mccarthyRecursiveFunctionsSymbolic1960}
\bibfield{author}{\bibinfo{person}{John McCarthy}.}
  \bibinfo{year}{1960}\natexlab{}.
\newblock \showarticletitle{Recursive {Functions} of {Symbolic} {Expressions}
  and {Their} {Computation} by {Machine}, {Part} {I}}.
\newblock \bibinfo{journal}{\emph{Commun. ACM}} \bibinfo{volume}{3},
  \bibinfo{number}{4} (\bibinfo{date}{April} \bibinfo{year}{1960}),
  \bibinfo{pages}{184--195}.
\newblock
\href{https://doi.org/10.1145/367177.367199}{doi:\nolinkurl{10.1145/367177.367199}}


\bibitem[{OpenJDK Contributors}(2024)]%
        {JDKversion}
\bibfield{author}{\bibinfo{person}{{OpenJDK Contributors}}.}
  \bibinfo{year}{2024}\natexlab{}.
\newblock \bibinfo{title}{{OpenJDK} {JDK} mainline, tag jdk-24+2, commit
  50bed6c}.
\newblock
\shownote{Commit 50bed6c67b1edd7736bdf79308d135a4e1047ff0}.
\newblock
\urldef\tempurl%
\url{https://github.com/openjdk/jdk/commit/50bed6c67b1edd7736bdf79308d135a4e1047ff0}
\showURL{%
\tempurl}


\bibitem[{Oracle Corporation}(2025)]%
        {oracle_tuning}
\bibfield{author}{\bibinfo{person}{{Oracle Corporation}}.}
  \bibinfo{year}{2025}\natexlab{}.
\newblock \bibinfo{title}{The {Z} {Garbage} {Collector}}.
\newblock
\shownote{Published: HotSpot Virtual Machine Garbage Collection Tuning Guide,
  Java SE 24}.
\newblock
\urldef\tempurl%
\url{https://docs.oracle.com/en/java/javase/24/gctuning/z-garbage-collector.html}
\showURL{%
\tempurl}


\bibitem[Shimchenko et~al\mbox{.}(2025)]%
        {shimchenko_monk_2025}
\bibfield{author}{\bibinfo{person}{Marina Shimchenko}, \bibinfo{person}{Erik
  Österlund}, {and} \bibinfo{person}{Tobias Wrigstad}.}
  \bibinfo{year}{2025}\natexlab{}.
\newblock \showarticletitle{Monk: {Opportunistic} {Scheduling} to {Delay}
  {Horizontal} {Scaling}}.
\newblock \bibinfo{journal}{\emph{The Art, Science, and Engineering of
  Programming}} \bibinfo{volume}{10}, \bibinfo{number}{1} (\bibinfo{date}{Feb.}
  \bibinfo{year}{2025}), \bibinfo{pages}{1}.
\newblock
\shownote{arXiv:2502.20522 [cs.PL]}.
\newblock
\showISSN{2473-7321}
\href{https://doi.org/10.22152/programming-journal.org/2026/10/1}{doi:\nolinkurl{10.22152/programming-journal.org/2026/10/1}}


\bibitem[{Standard Performance Evaluation Corporation}(2015)]%
        {specjbb}
\bibfield{author}{\bibinfo{person}{{Standard Performance Evaluation
  Corporation}}.} \bibinfo{year}{2015}\natexlab{}.
\newblock \bibinfo{title}{{SPECjbb}2015 {Benchmark}, version 1.03}.
\newblock
\shownote{Standard Performance Evaluation Corporation}.
\newblock
\urldef\tempurl%
\url{https://www.spec.org/jbb2015/}
\showURL{%
\tempurl}


\bibitem[{Stefan Karlsson}(2021)]%
        {stefan_karlsson_jep_2021}
\bibfield{author}{\bibinfo{person}{{Stefan Karlsson}}.}
  \bibinfo{year}{2021}\natexlab{}.
\newblock \bibinfo{booktitle}{\emph{{JEP} 439: {Generational} {ZGC}}}.
\newblock \bibinfo{type}{{T}echnical {R}eport}.
\newblock
\urldef\tempurl%
\url{https://openjdk.org/jeps/439}
\showURL{%
\tempurl}


\bibitem[Tavakolisomeh et~al\mbox{.}(2023)]%
        {tavakolisomeh_heap_2023}
\bibfield{author}{\bibinfo{person}{Sanaz Tavakolisomeh},
  \bibinfo{person}{Marina Shimchenko}, \bibinfo{person}{Erik Österlund},
  \bibinfo{person}{Rodrigo Bruno}, \bibinfo{person}{Paulo Ferreira}, {and}
  \bibinfo{person}{Tobias Wrigstad}.} \bibinfo{year}{2023}\natexlab{}.
\newblock \showarticletitle{Heap {Size} {Adjustment} with {CPU} {Control}}. In
  \bibinfo{booktitle}{\emph{Proceedings of the 20th {ACM} {SIGPLAN}
  {International} {Conference} on {Managed} {Programming} {Languages} and
  {Runtimes}}} \emph{(\bibinfo{series}{{MPLR} 2023})}.
  \bibinfo{publisher}{Association for Computing Machinery},
  \bibinfo{address}{New York, NY, USA}, \bibinfo{pages}{114--128}.
\newblock
\showISBNx{979-8-4007-0380-5}
\href{https://doi.org/10.1145/3617651.3622988}{doi:\nolinkurl{10.1145/3617651.3622988}}


\bibitem[White et~al\mbox{.}(2013)]%
        {whiteControlTheoryPrincipled2013}
\bibfield{author}{\bibinfo{person}{David~R. White}, \bibinfo{person}{Jeremy
  Singer}, \bibinfo{person}{Jonathan~M. Aitken}, {and}
  \bibinfo{person}{Richard~E. Jones}.} \bibinfo{year}{2013}\natexlab{}.
\newblock \showarticletitle{Control Theory for Principled Heap Sizing}. In
  \bibinfo{booktitle}{\emph{Proceedings of the 2013 International Symposium on
  Memory Management}}. \bibinfo{publisher}{ACM}, \bibinfo{address}{Seattle
  Washington USA}, \bibinfo{pages}{27--38}.
\newblock
\showISBNx{978-1-4503-2100-6}
\href{https://doi.org/10.1145/2464157.2466481}{doi:\nolinkurl{10.1145/2464157.2466481}}


\bibitem[Wilson(1992)]%
        {wilsonUniprocessorGarbageCollection}
\bibfield{author}{\bibinfo{person}{Paul~R. Wilson}.}
  \bibinfo{year}{1992}\natexlab{}.
\newblock \showarticletitle{Uniprocessor {Garbage} {Collection} {Techniques}}.
  In \bibinfo{booktitle}{\emph{Proceedings of the {International} {Workshop} on
  {Memory} {Management} ({IWMM} '92)}} \emph{(\bibinfo{series}{Lecture {Notes}
  in {Computer} {Science}}, Vol.~\bibinfo{volume}{637})}.
  \bibinfo{publisher}{Springer-Verlag}, \bibinfo{address}{Berlin, Heidelberg},
  \bibinfo{pages}{1--42}.
\newblock
\href{https://doi.org/10.1007/BFb0017182}{doi:\nolinkurl{10.1007/BFb0017182}}


\bibitem[Yang and Wrigstad(2022)]%
        {yang_deep_2022}
\bibfield{author}{\bibinfo{person}{Albert~Mingkun Yang} {and}
  \bibinfo{person}{Tobias Wrigstad}.} \bibinfo{year}{2022}\natexlab{}.
\newblock \showarticletitle{Deep {Dive} into {ZGC}: {A} {Modern} {Garbage}
  {Collector} in {OpenJDK}}.
\newblock \bibinfo{journal}{\emph{ACM Transactions on Programming Languages and
  Systems}} \bibinfo{volume}{44}, \bibinfo{number}{4} (\bibinfo{date}{Dec.}
  \bibinfo{year}{2022}), \bibinfo{pages}{1--34}.
\newblock
\showISSN{0164-0925, 1558-4593}
\href{https://doi.org/10.1145/3538532}{doi:\nolinkurl{10.1145/3538532}}


\bibitem[Yang et~al\mbox{.}(2004)]%
        {yangAutomaticHeapSizing2004}
\bibfield{author}{\bibinfo{person}{Ting Yang}, \bibinfo{person}{Emery~D.
  Berger}, \bibinfo{person}{Scott~F. Kaplan}, {and}
  \bibinfo{person}{J.~Eliot~B. Moss}.} \bibinfo{year}{2004}\natexlab{}.
\newblock \showarticletitle{Automatic {Heap} {Sizing}: {Taking} {Real} {Memory}
  into {Account}}. In \bibinfo{booktitle}{\emph{Proceedings of the 4th
  {International} {Symposium} on {Memory} {Management}}}
  \emph{(\bibinfo{series}{{ISMM} '04})}. \bibinfo{publisher}{Association for
  Computing Machinery}, \bibinfo{address}{New York, NY, USA},
  \bibinfo{pages}{61--72}.
\newblock
\href{https://doi.org/10.1145/1029873.1029881}{doi:\nolinkurl{10.1145/1029873.1029881}}


\bibitem[Yang et~al\mbox{.}(2006)]%
        {yang2006osdi}
\bibfield{author}{\bibinfo{person}{Ting Yang}, \bibinfo{person}{Emery~D.
  Berger}, \bibinfo{person}{Scott~F. Kaplan}, {and}
  \bibinfo{person}{J.~Eliot~B. Moss}.} \bibinfo{year}{2006}\natexlab{}.
\newblock \showarticletitle{{CRAMM}: Virtual Memory Support for
  {Garbage-Collected} Applications}. In \bibinfo{booktitle}{\emph{7th USENIX
  Symposium on Operating Systems Design and Implementation (OSDI 06)}}.
  \bibinfo{publisher}{USENIX Association}, \bibinfo{address}{Seattle, WA}.
\newblock
\urldef\tempurl%
\url{https://www.usenix.org/conference/osdi-06/cramm-virtual-memory-support-garbage-collected-applications}
\showURL{%
\tempurl}


\bibitem[Yuasa(1990)]%
        {yuasa1990}
\bibfield{author}{\bibinfo{person}{Taiichi Yuasa}.}
  \bibinfo{year}{1990}\natexlab{}.
\newblock \showarticletitle{Real-{Time} {Garbage} {Collection} on
  {General}-{Purpose} {Machines}}.
\newblock \bibinfo{journal}{\emph{Journal of Systems and Software}}
  \bibinfo{volume}{11}, \bibinfo{number}{3} (\bibinfo{year}{1990}),
  \bibinfo{pages}{181--198}.
\newblock
\href{https://doi.org/10.1016/0164-1212(90)90084-Y}{doi:\nolinkurl{10.1016/0164-1212(90)90084-Y}}


\bibitem[Zhao et~al\mbox{.}(2022)]%
        {zhaoImprovingConcurrentGC2022}
\bibfield{author}{\bibinfo{person}{Junxian Zhao}, \bibinfo{person}{Aidi Pi},
  \bibinfo{person}{Xiaobo Zhou}, \bibinfo{person}{Sang-Yoon Chang}, {and}
  \bibinfo{person}{Chengzhong Xu}.} \bibinfo{year}{2022}\natexlab{}.
\newblock \showarticletitle{Improving {Concurrent} {GC} for {Latency}
  {Critical} {Services} in {Multi}-tenant {Systems}}. In
  \bibinfo{booktitle}{\emph{Proceedings of the 23rd {ACM}/{IFIP}
  {International} {Middleware} {Conference}}}
  \emph{(\bibinfo{series}{{Middleware} '22})}. \bibinfo{publisher}{Association
  for Computing Machinery}, \bibinfo{address}{New York, NY, USA},
  \bibinfo{pages}{43--55}.
\newblock
\href{https://doi.org/10.1145/3528535.3531515}{doi:\nolinkurl{10.1145/3528535.3531515}}


\bibitem[Österlund(2024)]%
        {automatedHeapSizingTalk}
\bibfield{author}{\bibinfo{person}{Erik Österlund}.}
  \bibinfo{year}{2024}\natexlab{}.
\newblock \bibinfo{title}{{ZGC} {Automatic} {Heap} {Sizing}}.
\newblock
\shownote{Presented at the JVM Language Summit (JVMLS)}.
\newblock
\urldef\tempurl%
\url{https://inside.java/2024/11/09/jvmls-zgc/}
\showURL{%
\tempurl}


\bibitem[Österlund(2026)]%
        {osterlund_jep_2026}
\bibfield{author}{\bibinfo{person}{Erik Österlund}.}
  \bibinfo{year}{2026}\natexlab{}.
\newblock \bibinfo{booktitle}{\emph{{JEP} {Draft} 8377305: {Automatic} {Heap}
  {Sizing} for {ZGC}}}.
\newblock \bibinfo{type}{{T}echnical {R}eport}.
\newblock
\shownote{Published: OpenJDK}.
\newblock
\urldef\tempurl%
\url{https://openjdk.org/jeps/8377305}
\showURL{%
\tempurl}


\end{thebibliography}

\clearpage
\appendix
\section{Young and Old Collection Cycles for ZGC}
\label{sec:zgc_cycles}
\begin{figure}
    \centering
    \includegraphics[width=\linewidth]{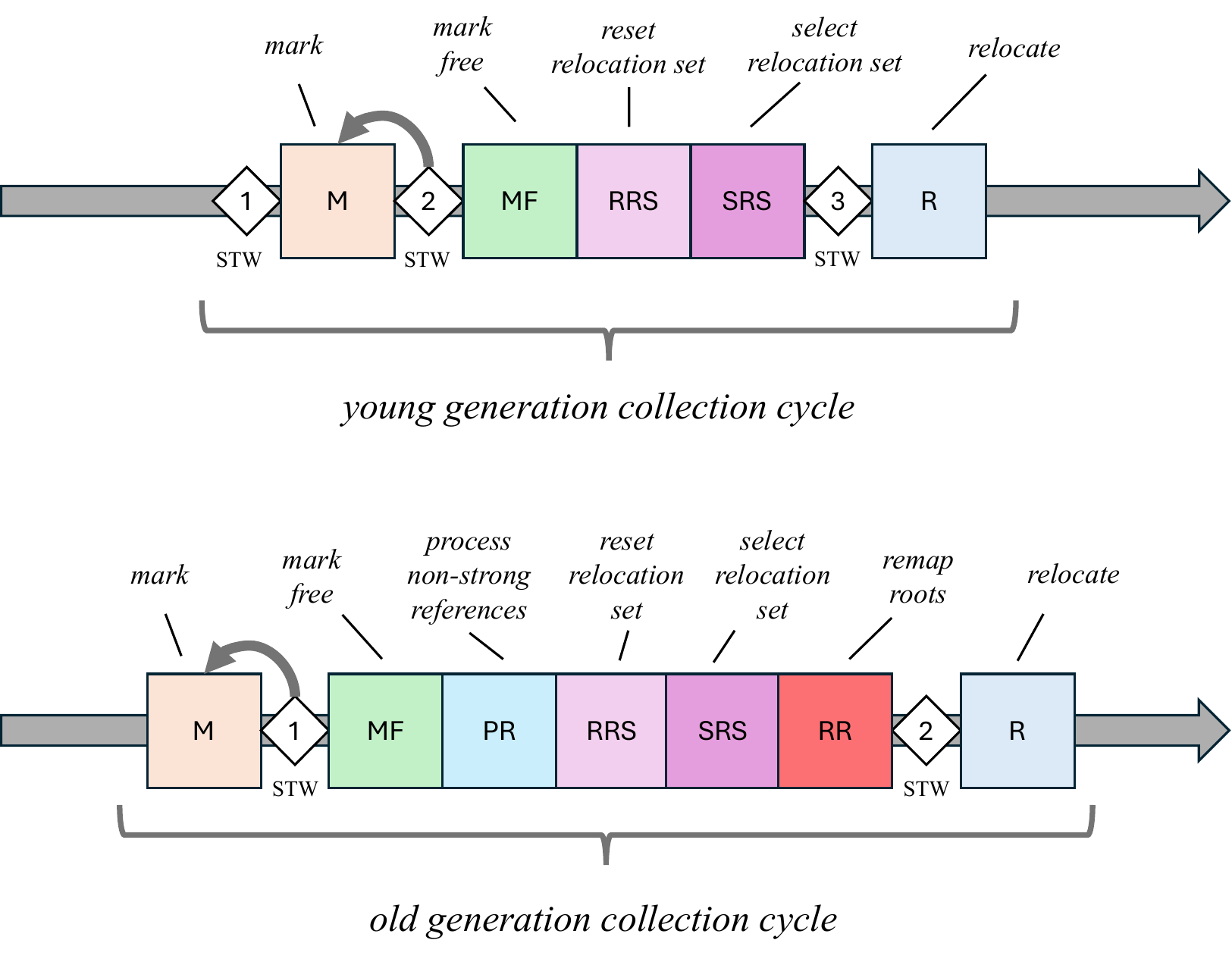}
    \caption{Young and old collection cycles in ZGC. Old collection does not require an initial pause because it always runs as part of a major GC, immediately after young collection. Box widths are not proportional to phase lengths.}
    \Description{Flowchart showing main phases in the young and old collection cycles for ZGC.}
    \label{fig:zgc_cycles}
\end{figure}

Figure~\ref{fig:zgc_cycles} illustrates the young and old collection cycles in ZGC.
Young generation collection consists of three brief stop-the-world (STW) pauses and five concurrent phases: mark (M), mark free (MF), reset relocation set (RRS), select relocation set (SRS), and relocate (R).
The initial pause flips the global status bits so that the barrier instructions in the mutators see that a new marking phase has begun and also scans the mutators' stacks and other roots to seed the mark.
Next, the mark phase traverses the young object graph from the roots, marking any objects it reaches.
The second pause attempts to terminate the mark, but a concurrent mutator might have created additional work during marking (e.g., by writing an object reference) that is not yet complete.
In that case, the cycle retries the mark phase until it completes with no additional work.
When marking is complete, the mark free phase releases some per-worker marking data structures.
Next, the cycle prepares for object relocation by: 1) clearing the previous cycle's relocation structures, and 2) scanning the candidate ZPages and selecting the sparse ZPages from which it will evacuate live objects.
The third pause sets the global status bits to indicate the beginning of the relocation phase and relocates root objects so that mutators immediately see the forwarded root references.
The relocation phase then proceeds concurrently with the mutators, moving the live objects from the selected ZPages to new ZPages and freeing the sparse (now unused) ZPages.
Note that, during this process, if a mutator attempts to use an object that is marked for relocation but has not yet been moved, the mutator itself will move the object via the load barrier.

The old collection cycle largely mirrors the young cycle, but with one fewer pause and two additional phases.
Since each old collection is always immediately preceded by a young collection, the synchronization steps that are necessary prior to the mark phase, including updating the global status bits, can be completed during the young cycle's initial pause.
Hence, the old cycle omits the initial STW pause.
The extra phases in the old cycle are process non-strong references (PR) and remap roots (RR).
Java has several types of ``non-strong'' references (e.g., soft and weak references) whose referents are reclaimed when they are no longer strongly reachable across the entire heap.
The PR phase resolves these non-strong references using the strong reachability information established during marking.
Finally, the remap roots phase heals stale pointers left by previous cycles' relocations before the global relocation bits flip again.
In this way, mutators avoid misinterpreting pointers that are stale across multiple relocation cycles.

\clearpage
%
%

\newcommand{\appwide}[3]{%
  {\centering\includegraphics[width=0.98\textwidth]{#1}\par}%
  \captionof{figure}{#2}\label{#3}\vspace{1.5ex}}
\newcommand{\apphalf}[3]{%
  \begin{minipage}[t]{0.49\textwidth}%
    \centering\includegraphics[width=\textwidth]{#1}%
    \captionof{figure}{#2}\label{#3}%
  \end{minipage}}
\newcommand{\appwidep}[4]{%
  {\centering\includegraphics[page=#1,width=0.80\textwidth]{#2}\par}%
  \captionof{figure}{#3}\label{#4}\vspace{1.5ex}}
\newcommand{\apphalfp}[4]{%
  \begin{minipage}[t]{0.49\textwidth}%
    \centering\includegraphics[page=#1,width=\textwidth]{#2}%
    \captionof{figure}{#3}\label{#4}%
  \end{minipage}}
\newcommand{\approwsep}{\vspace{1.5ex}}

\twocolumn[%
  \section{DaCapo Benchmark Characteristics}\label{sec:dacapo_table}
  {\captionsetup{type=table}\centering\small
\begin{tabular}{|c|c|r|r|r|r||r|r|r|}
\hline
\multirow{2}{*}{Benchmark} & \multirow{2}{*}{Iters} &
\multicolumn{1}{c|}{Min} &
\multicolumn{3}{c||}{Baseline ZGC with -Xmx32g} &
\multicolumn{3}{c|}{Baseline ZGC with -Xmx160g} \\
\cline{4-9}
 & & \multicolumn{1}{c|}{MB} &
 \multicolumn{1}{c|}{Time (s)} &
 \multicolumn{1}{c|}{CPU \%} &
 \multicolumn{1}{c||}{Max GB} &
 \multicolumn{1}{c|}{Time (s)} &
 \multicolumn{1}{c|}{CPU \%} &
 \multicolumn{1}{c|}{Max GB} \\
\hline
\multicolumn{9}{|c|}{\textbf{Default Input}} \\
\hline
cassandra$^\dagger$ & 10  & 182  & 67.6   & 188.6   & 25.6 &  67.7  & 189.6   & 29.2 \\
fop                 & 100 & 26   & 60.8   & 199.2   & 17.6 &  62.4  & 187.0   & 21.6 \\
graphchi            & 10  & 196  & 54.7   & 312.2   & 25.6 &  58.2  & 331.8   & 48.0 \\
h2$^\dagger$        & 10  & 980  & 37.5   & 452.2   & 18.1 &  42.4  & 418.4   & 108.4 \\
jython              & 10  & 58   & 64.3   & 154.8   & 15.1 &  71.0  & 140.6   & 27.1 \\
lusearch$^\dagger$  & 10  & 42   & 129.2  & 1,431.8 & 24.1 &  134.5 & 1,422.8 & 48.0 \\
pmd                 & 30  & 216  & 45.6   & 628.4   & 24.0 &  58.0  & 653.0   & 104.7 \\
spring$^\dagger$    & 30  & 98   & 104.8  & 1,209.2 & 25.6 &  113.3 & 1,153.6 & 143.3 \\
sunflow             & 30  & 50   & 118.6  & 1,501.0 & 25.7 &  106.9 & 1,441.6 & 60.2 \\
tomcat$^\dagger$    & 20  & 46   & 149.0  & 318.4   & 20.3 &  149.4 & 314.2   & 48.0 \\
xalan               & 50  & 40   & 42.1   & 1,363.6 & 25.7 &  49.7  & 1,260.0 & 65.4 \\
zxing               & 50  & 98   & 78.4   & 1,265.6 & 25.7 &  85.0  & 1,278.8 & 34.5 \\
\hline
\multicolumn{9}{|c|}{\textbf{Large Input}} \\
\hline
cassandra$^\dagger$ & 5 & 270    & 116.3   & 471.8   & 20.9 & 126.5   & 456.0   & 115.7 \\
graphchi            & 5 & 1,166  & 1,041.1 & 493.0   & 18.0 & 1,036.2 & 493.0   & 48.0 \\
h2$^\dagger$        & 5 & 14,230 & 710.3   & 727.4   & 26.2 & 692.9   & 688.0   & 115.5 \\
jython              & 5 & 60     & 620.3   & 123.2   & 23.3 & 660.1   & 119.4   & 117.0 \\
lusearch$^\dagger$  & 5 & 118    & 2,685.9 & 1,577.6 & 7.7  & 2,749.7 & 1,574.0 & 48.0 \\
pmd                 & 5 & 4,544  & 76.5    & 897.2   & 28.5 & 76.5    & 798.8   & 102.2 \\
spring$^\dagger$    & 5 & 134    & 306.0   & 1,298.4 & 25.6 & 313.1   & 1,258.0 & 152.0 \\
sunflow             & 5 & 258    & 267.7   & 1,550.8 & 25.7 & 262.4   & 1,533.0 & 58.4 \\
tomcat$^\dagger$    & 5 & 56     & 354.3   & 293.0   & 24.3 & 355.1   & 292.4   & 48.0 \\
xalan               & 5 & 52     & 67.2    & 1,342.4 & 25.7 & 73.0    & 1,248.6 & 85.9 \\
\hline
\end{tabular}
  \caption{DaCapo benchmarks with baseline performance characteristics, grouped by input size. In the first column, $^\dagger$ marks benchmarks that report request latency after each iteration. The next two columns show the \# of iterations and the minimum heap size in MB. The next three columns show the total time to run the benchmark, CPU utilization as a \% of all computing cores (as reported by \texttt{time}, out of 1,600\%), and the maximum heap usage (in GB) in the baseline configuration with -Xmx32g. The final three columns show the same measurements for the baseline with -Xmx set to 160 GB. All measurements were collected on the Intel Xeon platform.}
  \label{tab:dacapo}}
]

\clearpage

\twocolumn[%
  \section{Supplemental Results}
  \label{sec:supplemental}
  \subsection{Intel Xeon Platform}
  \subsubsection{Hard Max and Soft Max with Varying CPU Constraints}\label{sec:core_constraints}\leavevmode\par\vspace{0.5ex}
  {\captionsetup{type=figure}\centering
  \begin{minipage}[t]{0.49\textwidth}\centering
    \includegraphics[width=\linewidth]{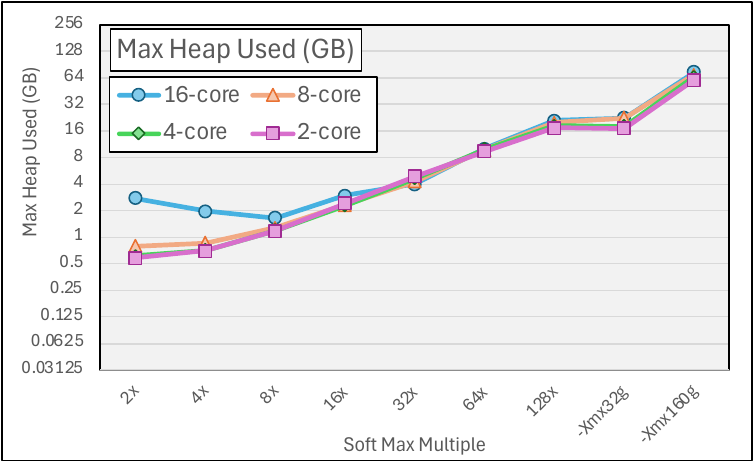}
    \subcaption{Max heap used (GB, log scale).}\label{fig:core_max_used}
  \end{minipage}\hfill
  \begin{minipage}[t]{0.49\textwidth}\centering
    \includegraphics[width=\linewidth]{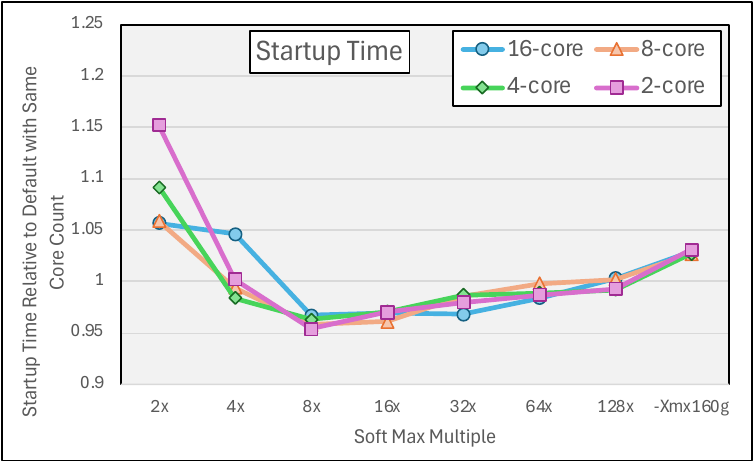}
    \subcaption{Startup time.}\label{fig:core_startup}
  \end{minipage}

  \vspace{1.5ex}

  \begin{minipage}[t]{0.49\textwidth}\centering
    \includegraphics[width=\linewidth]{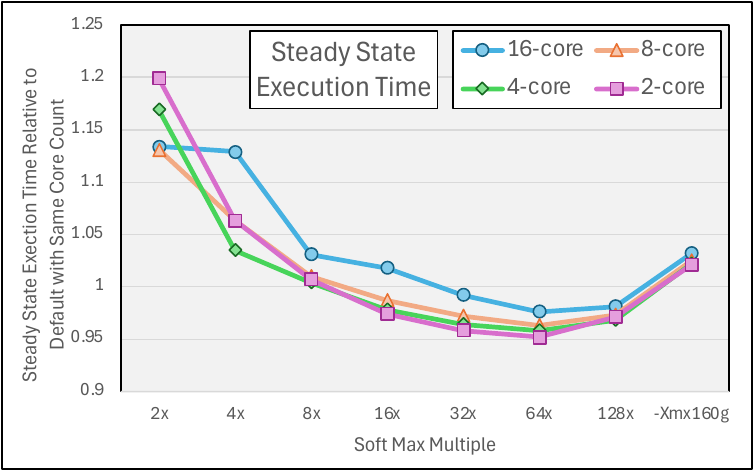}
    \subcaption{Steady-state execution time.}\label{fig:core_steady_state}
  \end{minipage}\hfill
  \begin{minipage}[t]{0.49\textwidth}\centering
    \includegraphics[width=\linewidth]{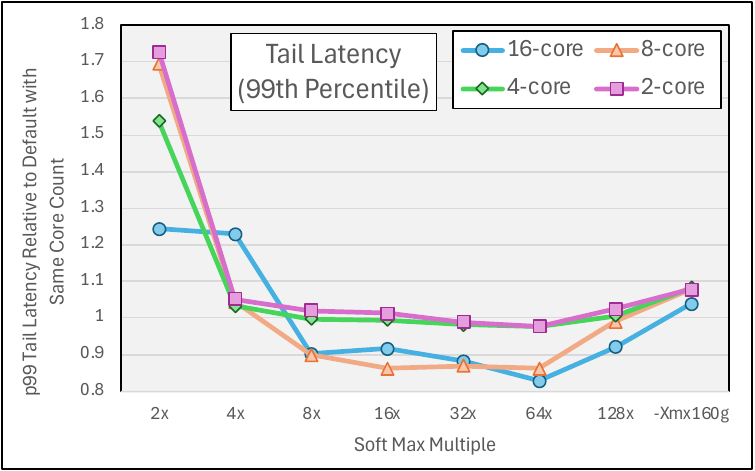}
    \subcaption{Tail latency (99th percentile).}\label{fig:core_latency}
      \end{minipage}
   \caption{Average results for the DaCapo benchmarks with \texttt{-XX:SoftMaxHeapSize} set to different multiples of the empirically-derived minimum heap size, with 2, 4, 8, or 16 cores available to the JVM. Performance results are shown relative to the default configuration with the same core count.}
  \label{fig:core_soft_max}}
]

\twocolumn[%
  \subsubsection{Per Benchmark Results for Evaluation with 16 Cores}\leavevmode\par\vspace{0.5ex}
  \label{sec:eval_16core_perbench} 
  \appwide{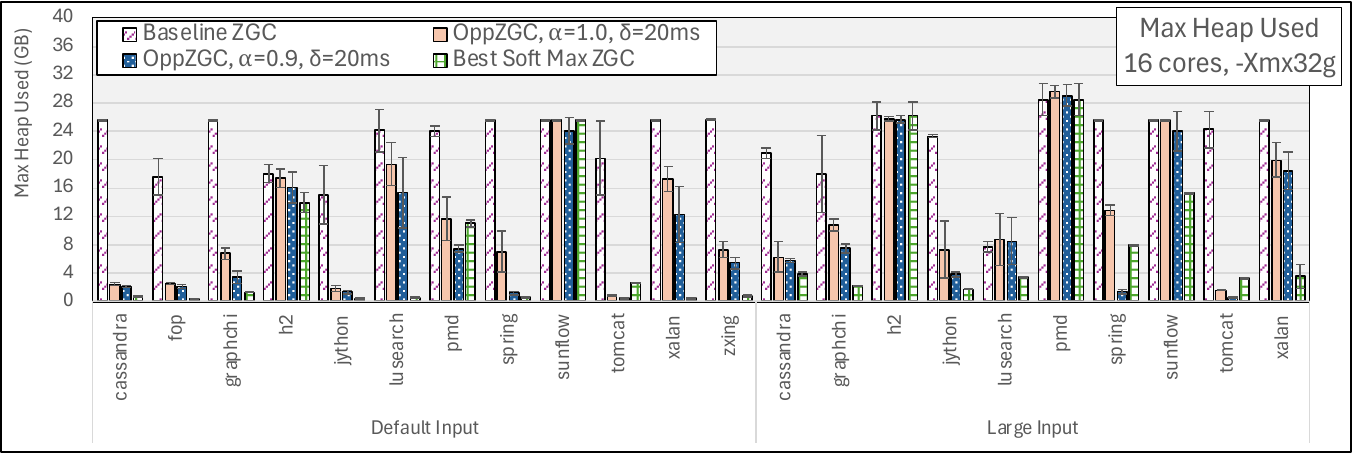}%
    {Maximum heap occupancy (GB) for each DaCapo benchmark, 16 cores, \texttt{-Xmx32g}.}{fig:app-maxheap-32g}
  \apphalf{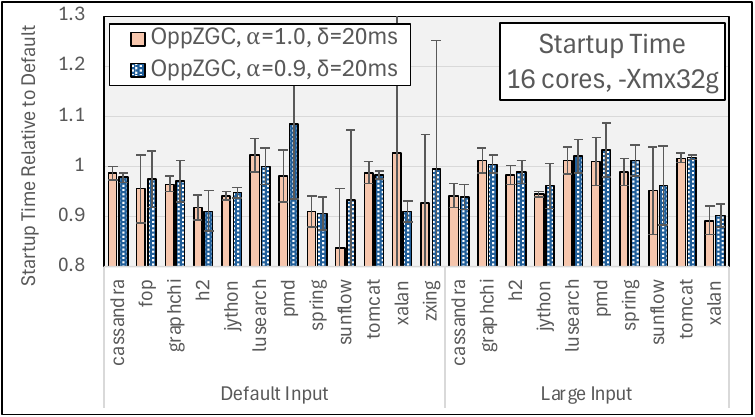}%
    {Startup time relative to baseline ZGC, 16 cores, \texttt{-Xmx32g}.}{fig:app-startup-32g}\hfill
  \apphalf{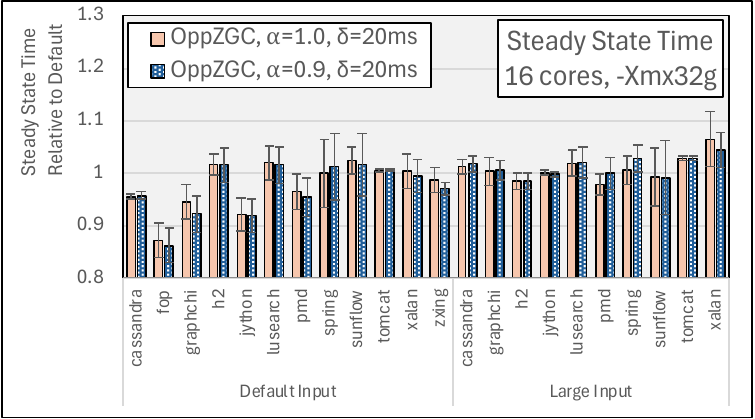}%
    {Steady-state time relative to baseline ZGC, 16 cores, \texttt{-Xmx32g}.}{fig:app-steady-32g}
  \approwsep
  \apphalf{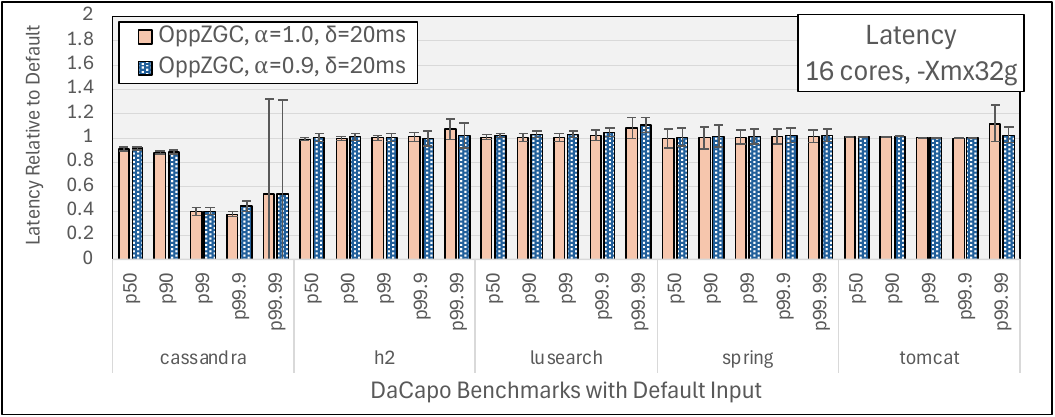}%
    {Latency distribution relative to baseline ZGC, default inputs, 16 cores, \texttt{-Xmx32g}.}{fig:app-latdef-32g}\hfill
  \apphalf{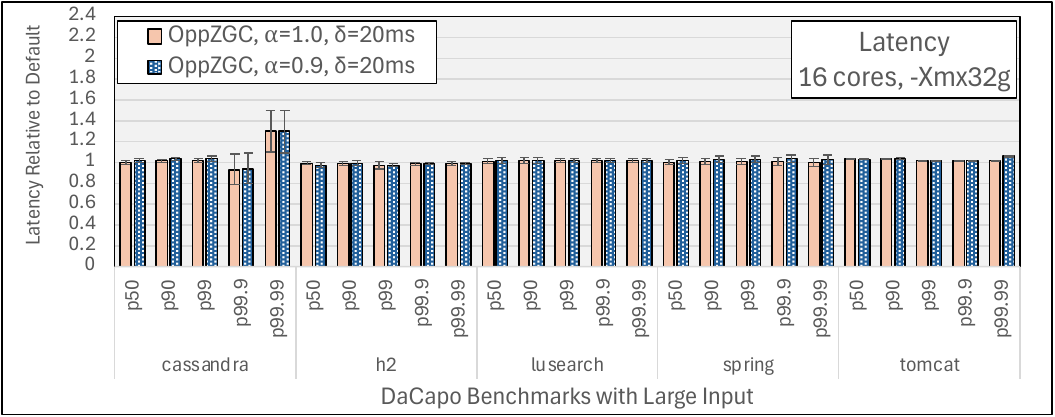}%
    {Latency distribution relative to baseline ZGC, large inputs, 16 cores, \texttt{-Xmx32g}.}{fig:app-latlarge-32g}
]

\twocolumn[%
  \appwide{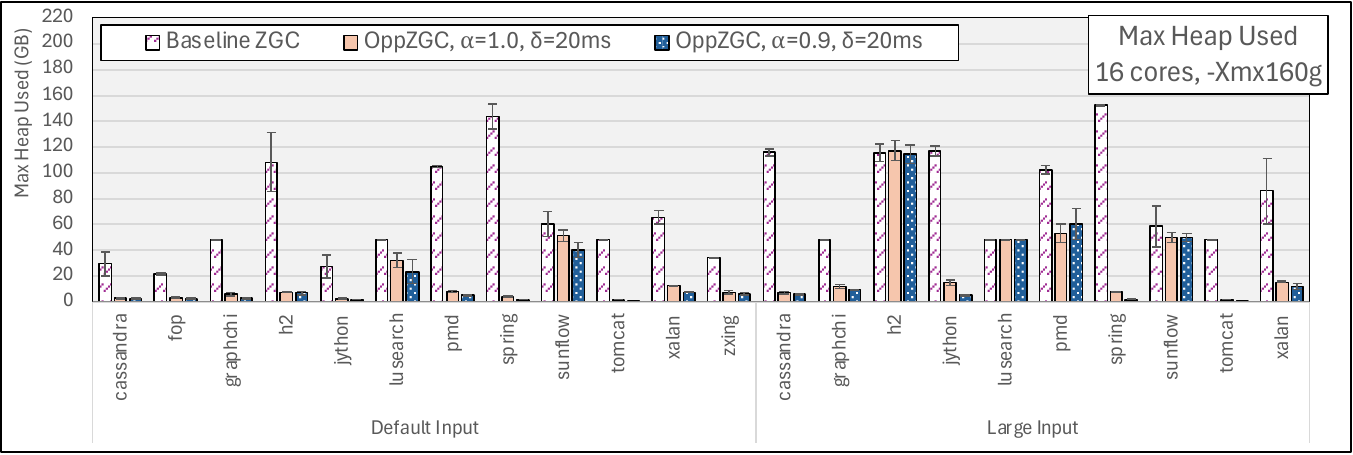}%
    {Maximum heap occupancy (GB) for each DaCapo benchmark, 16 cores, \texttt{-Xmx160g}.}{fig:app-maxheap-160g}
  \apphalf{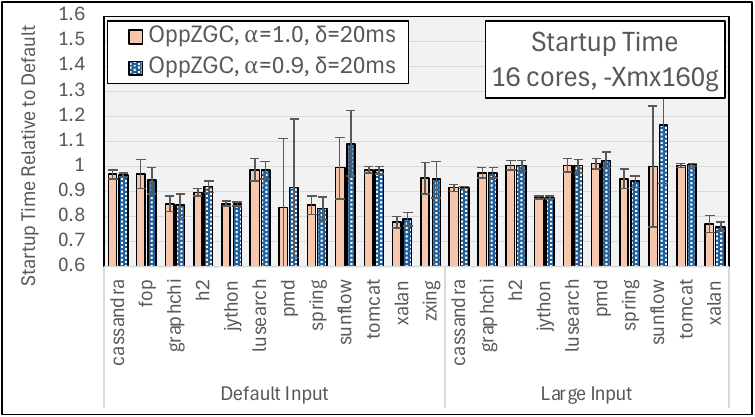}%
    {Startup time relative to baseline ZGC, 16 cores, \texttt{-Xmx160g}.}{fig:app-startup-160g}\hfill
  \apphalf{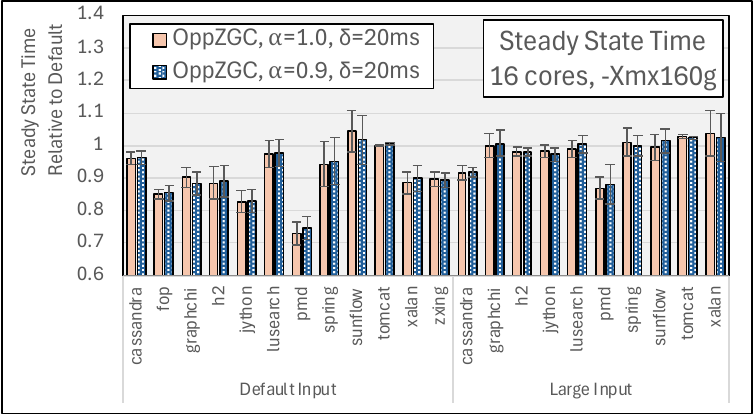}%
    {Steady-state time relative to baseline ZGC, 16 cores, \texttt{-Xmx160g}.}{fig:app-steady-160g}
      \approwsep
  \apphalf{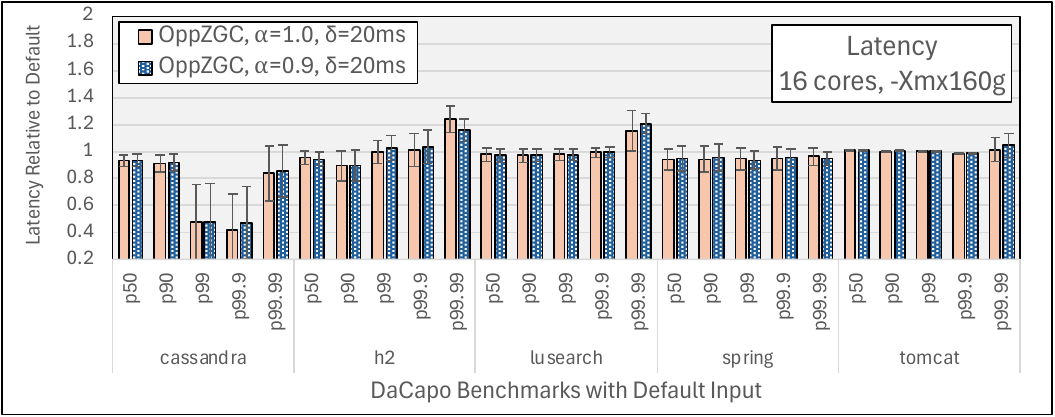}%
    {Latency distribution relative to baseline ZGC, default inputs, 16 cores, \texttt{-Xmx160g}.}{fig:app-latdef-160g}\hfill
  \apphalf{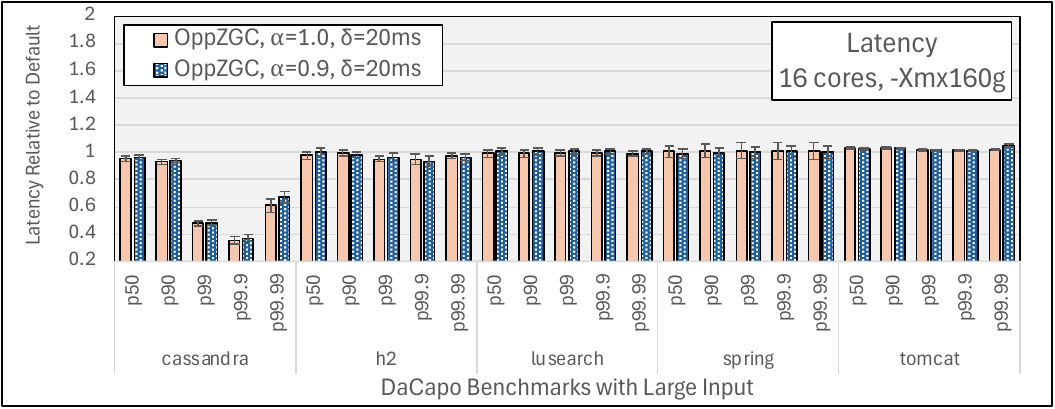}%
    {Latency distribution relative to baseline ZGC, large inputs, 16 cores, \texttt{-Xmx160g}.}{fig:app-latlarge-160g}
]

\twocolumn[%
  \subsubsection{Per Benchmark Results for Evaluation with Varying Cores}
  \label{sec:eval_varying_cores} 
  \leavevmode\par\vspace{0.5ex}
  \appwide{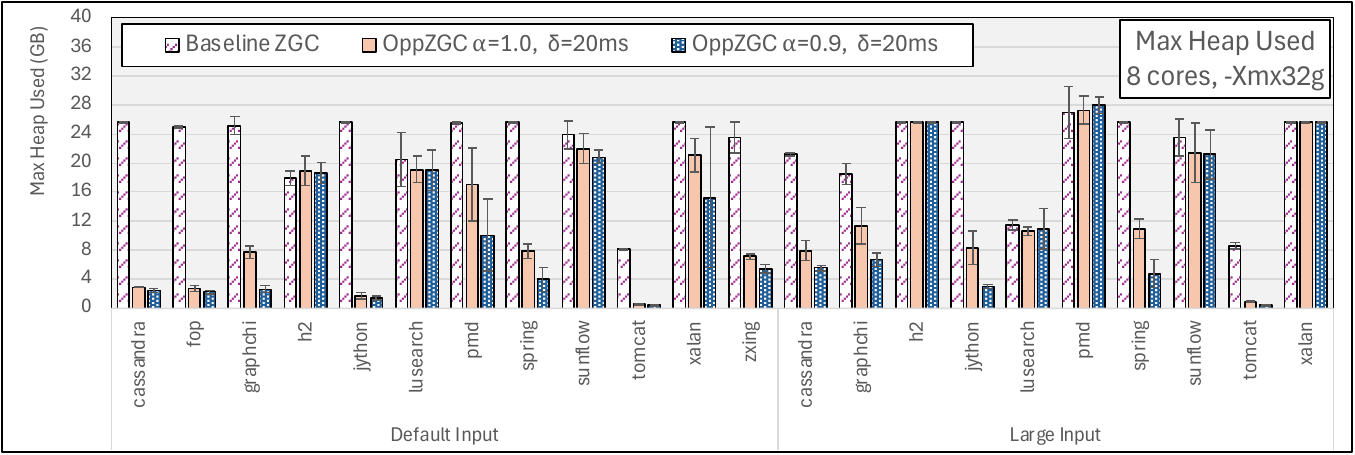}%
    {Maximum heap occupancy (GB) for each DaCapo benchmark, 8 cores, \texttt{-Xmx32g}.}{fig:app-maxheap-8c}
  \apphalf{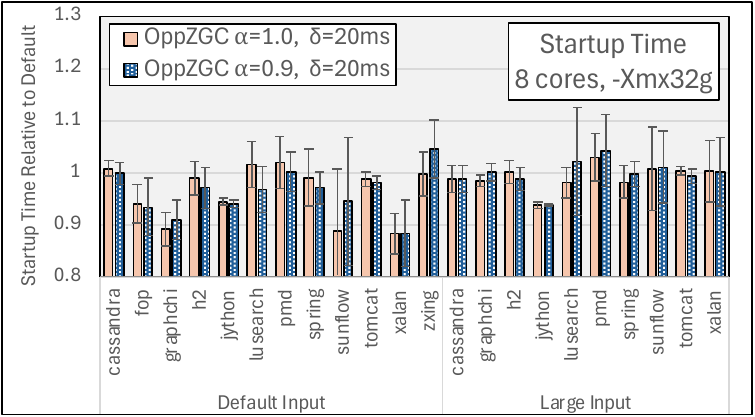}%
    {Startup time relative to baseline ZGC, 8 cores, \texttt{-Xmx32g}.}{fig:app-startup-8c}\hfill
  \apphalf{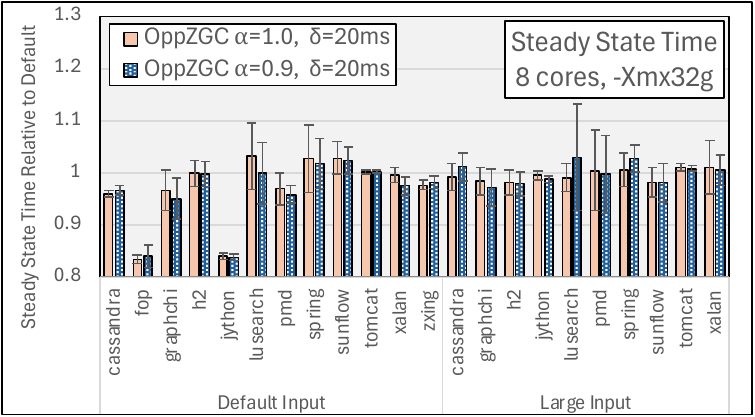}%
    {Steady-state time relative to baseline ZGC, 8 cores, \texttt{-Xmx32g}.}{fig:app-steady-8c}
  \approwsep
  \apphalf{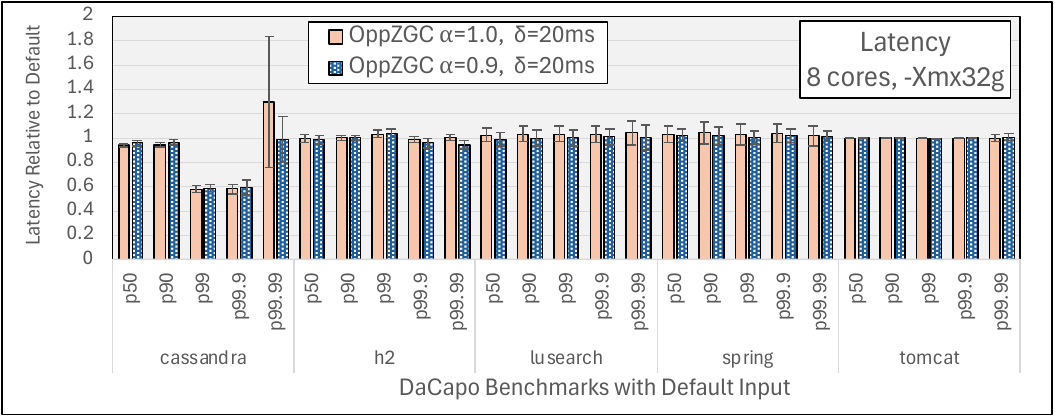}%
    {Latency distribution relative to baseline ZGC, default inputs, 8 cores, \texttt{-Xmx32g}.}{fig:app-latdef-8c}\hfill
  \apphalf{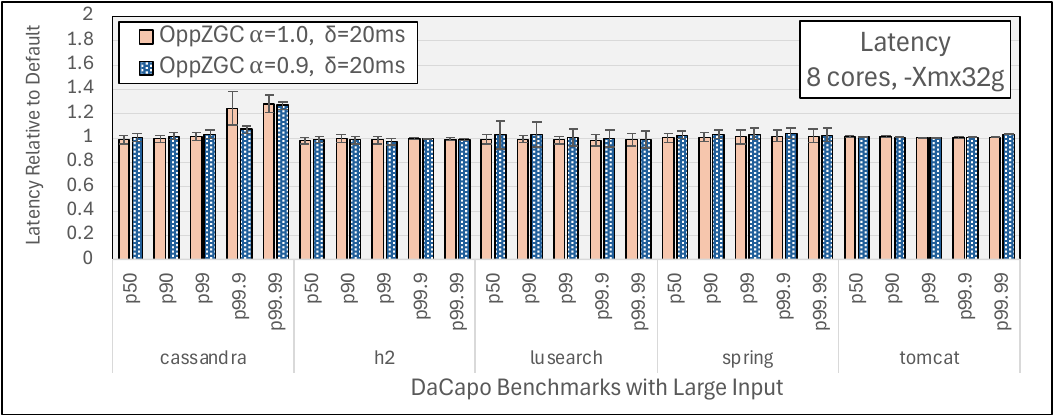}%
    {Latency distribution relative to baseline ZGC, large inputs, 8 cores, \texttt{-Xmx32g}.}{fig:app-latlarge-8c}
]

\twocolumn[%
  \appwide{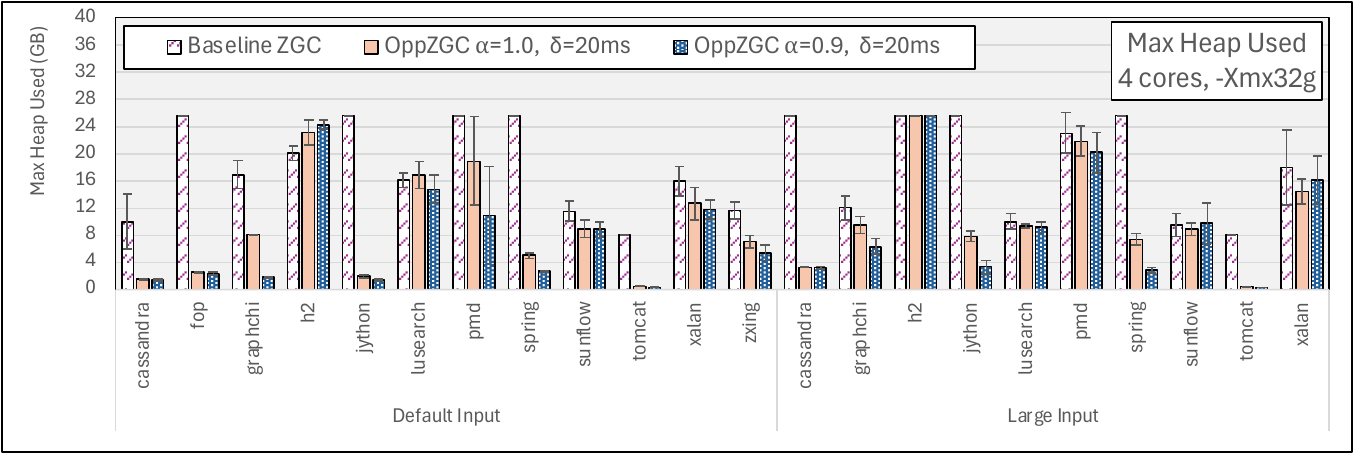}%
    {Maximum heap occupancy (GB) for each DaCapo benchmark, 4 cores, \texttt{-Xmx32g}.}{fig:app-maxheap-4c}
  \apphalf{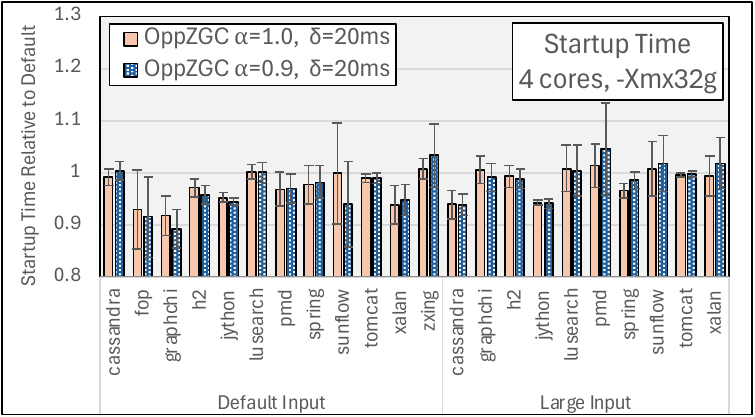}%
    {Startup time relative to baseline ZGC, 4 cores, \texttt{-Xmx32g}.}{fig:app-startup-4c}\hfill
  \apphalf{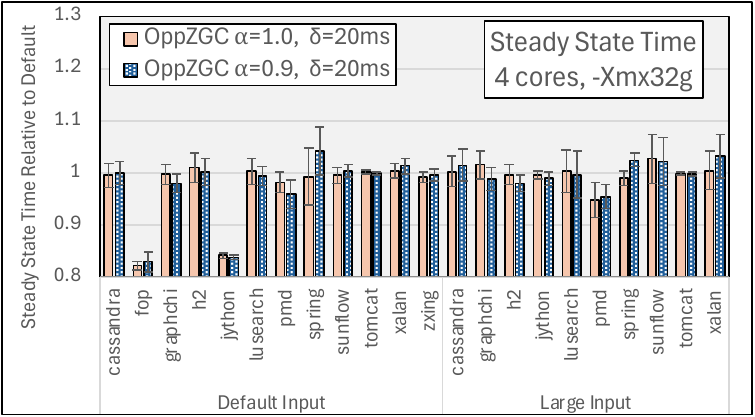}%
    {Steady-state time relative to baseline ZGC, 4 cores, \texttt{-Xmx32g}.}{fig:app-steady-4c}
  \approwsep
  \apphalf{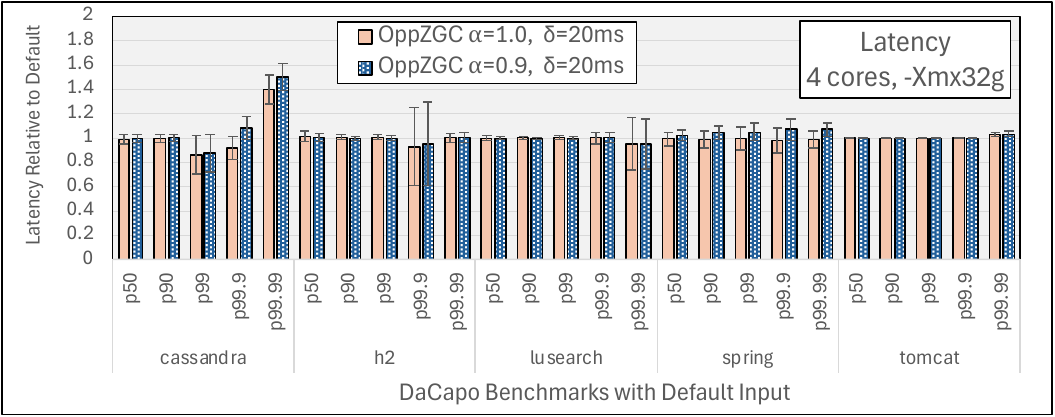}%
    {Latency distribution relative to baseline ZGC, default inputs, 4 cores, \texttt{-Xmx32g}.}{fig:app-latdef-4c}\hfill
  \apphalf{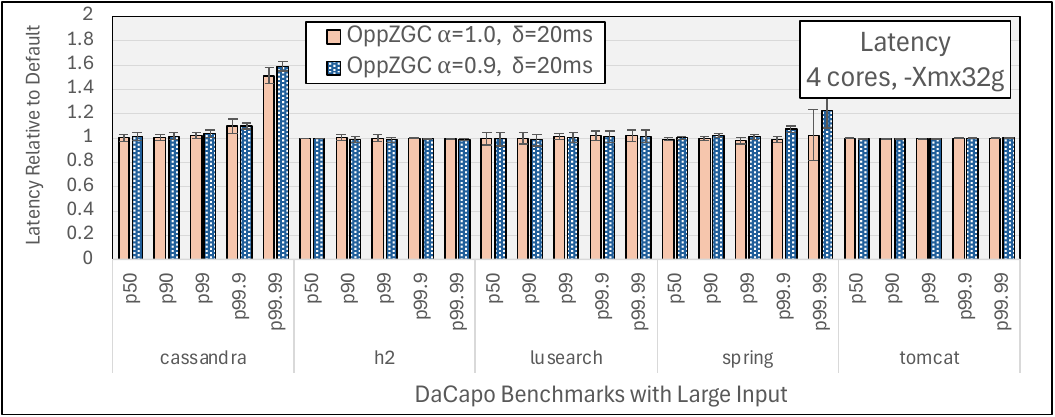}%
    {Latency distribution relative to baseline ZGC, large inputs, 4 cores, \texttt{-Xmx32g}.}{fig:app-latlarge-4c}
]

\twocolumn[%
  \appwide{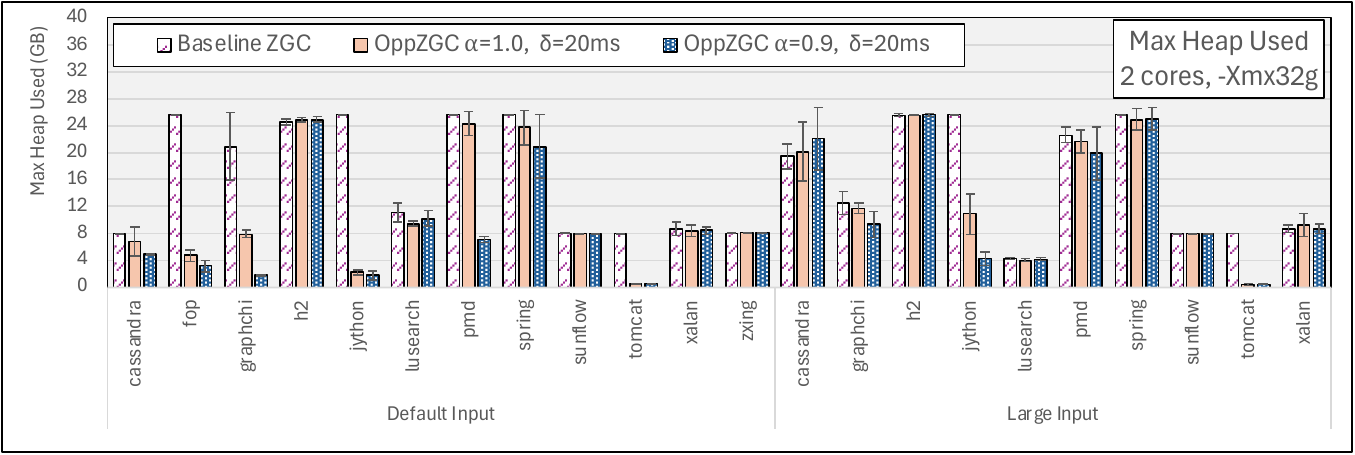}%
    {Maximum heap occupancy (GB) for each DaCapo benchmark, 2 cores, \texttt{-Xmx32g}.}{fig:app-maxheap-2c}
  \apphalf{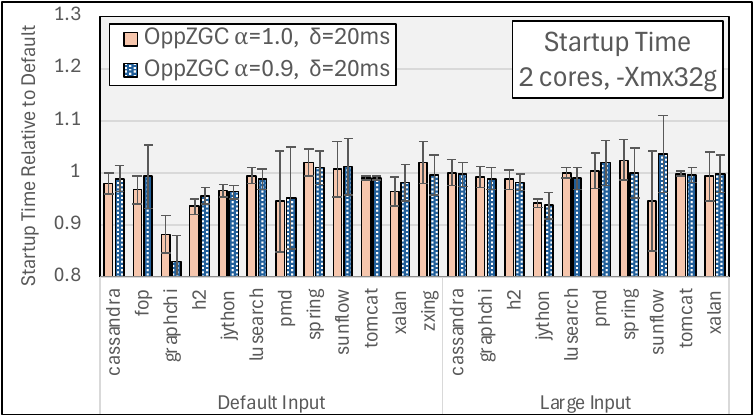}%
    {Startup time relative to baseline ZGC, 2 cores, \texttt{-Xmx32g}.}{fig:app-startup-2c}\hfill
  \apphalf{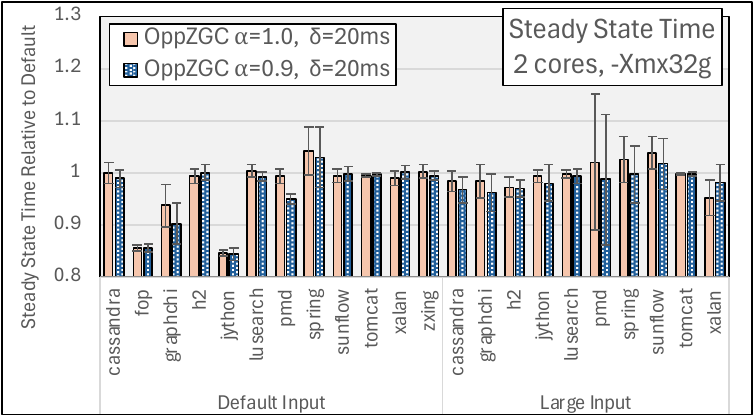}%
    {Steady-state time relative to baseline ZGC, 2 cores, \texttt{-Xmx32g}.}{fig:app-steady-2c}
  \approwsep
  \apphalf{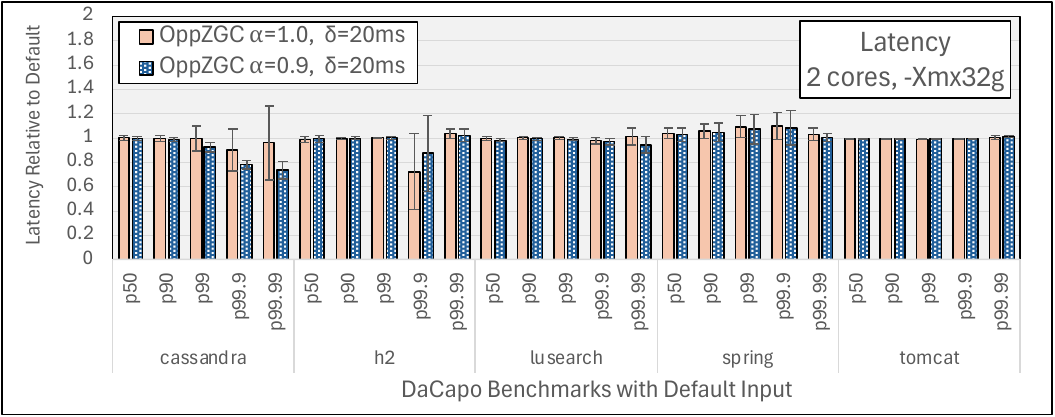}%
    {Latency distribution relative to baseline ZGC, default inputs, 2 cores, \texttt{-Xmx32g}.}{fig:app-latdef-2c}\hfill
  \apphalf{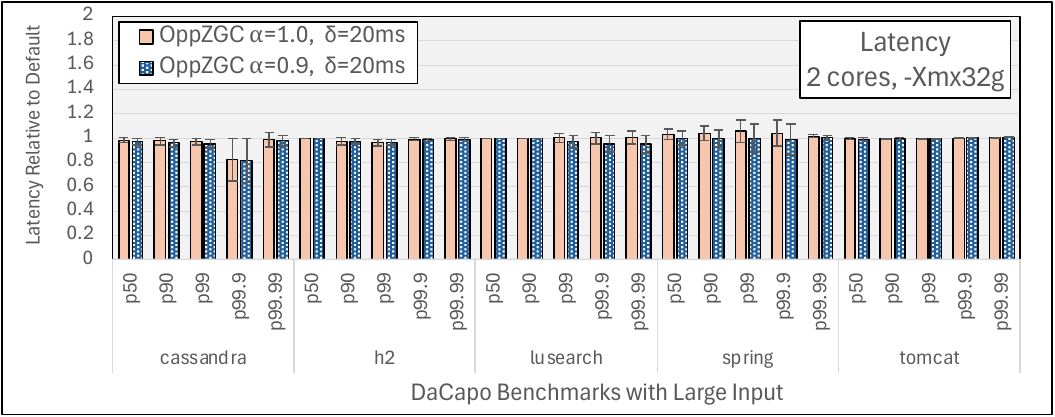}%
    {Latency distribution relative to baseline ZGC, large inputs, 2 cores, \texttt{-Xmx32g}.}{fig:app-latlarge-2c}
]

\clearpage

\subsection{AMD Platform}
\label{sec:amd_results}
\subsubsection{Platform Description}
Our second experimental platform contains a single AMD Ryzen~9 5950X CPU with 16 physical compute cores and simultaneous multithreading disabled.
The cores run at a 3.4~GHz base clock (boost disabled) and share a 64~MB L3 cache, split evenly across two 8-core core complex dies (CCDs), each with its own 32~MB L3 slice.
The processor's memory controller services two channels, each connected to two 16~GB, 3600~MT/s, DDR4 DIMMs, for a total of 64~GB of DDR4 SDRAM.
As such, all reported benchmarks on the AMD platform use \texttt{-Xmx32g}.

\subsubsection{Experimental Setup}
The evaluation on the AMD platform uses the same system and runtime software versions and configuration, as well as the same experimental configuration and parameters, as the evaluation on the Intel platform.
For the results in this appendix, we ran our DaCapo default and large input benchmarks and the SPECjbb2015 benchmark with three configurations: 1) baseline ZGC, 2) OppZGC with $\alpha=1.0$ and $\delta=20\,\mathrm{ms}$, and 3) OppZGC with $\alpha=0.9$ and $\delta=20\,\mathrm{ms}$.
All results reported here use all 16 cores of the platform and a maximum heap size of 32 GB.

\subsubsection{Evaluation Notes}
The results on the AMD platform show very similar performance and memory usage trends to the evaluation on the Intel platform.
On average, OppZGC reduces the maximum heap occupancy for the DaCapo default and large inputs by 59\% and 35\% with $\alpha=1.0$, and by 69\% and 45\% with $\alpha=0.9$.
Additionally, OppZGC exhibits similar startup time, throughput, and request latency to baseline ZGC with the DaCapo benchmarks, in almost all cases.
Notably, OppZGC increases p99.99 tail latency for \emph{cassandra} by about 84\% to 93\% with its large input, and by 36\% to 55\% with its default input, while leaving p99 and p99.9 at or below baseline.
For SPECjbb, we find OppZGC reduces the maximum heap occupancy by 20\%, while maintaining the same max jOPS and critical jOPS as baseline ZGC.
We are currently collecting core-restricted results for the DaCapo benchmarks on the AMD platform and plan to include these in the final version of this appendix.

\clearpage

\twocolumn[%
  \appwidep{3}{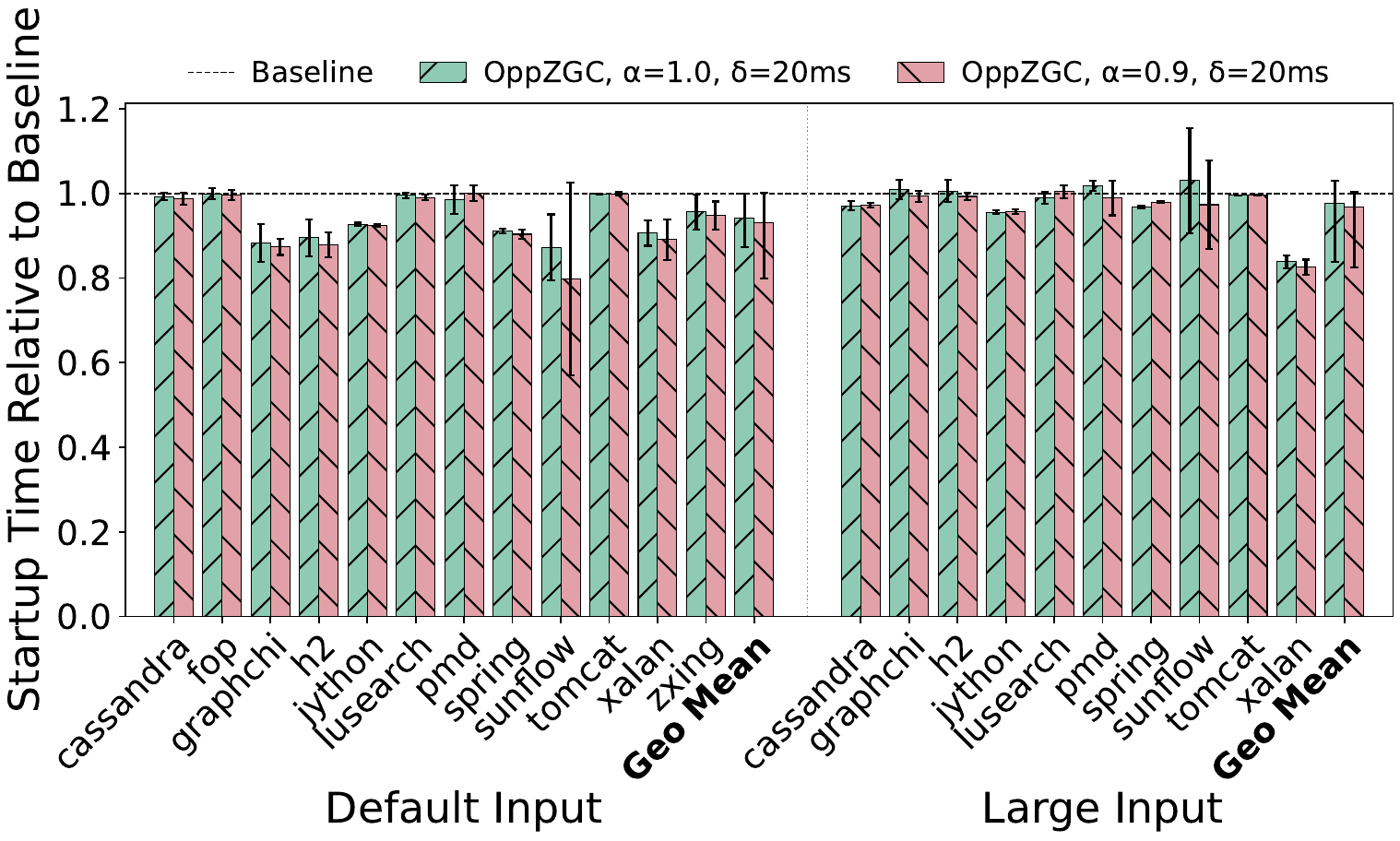}%
    {Maximum heap occupancy (in GB) under OppZGC, relative to baseline ZGC.}{fig:amd-dacapo-heap}
  \apphalfp{1}{figures/amd.pdf}%
    {Startup time.}{fig:amd-dacapo-startup}\hfill
  \apphalfp{2}{figures/amd.pdf}%
    {Steady-state time.}{fig:amd-dacapo-steady}
  \approwsep
  \apphalfp{5}{figures/amd.pdf}%
    {Latency distribution, default input.}{fig:amd-dacapo-tail-default}\hfill
  \apphalfp{6}{figures/amd.pdf}%
    {Latency distribution, large input.}{fig:amd-dacapo-tail-large}
]

\twocolumn[%
  \subsubsection{SPECjbb2015 Results on the AMD Platform}\leavevmode\par\vspace{0.5ex}
  \apphalfp{9}{figures/amd.pdf}%
    {Throughput.}{fig:amd-specjbb-jops}\hfill
  \apphalfp{10}{figures/amd.pdf}%
    {Memory usage.}{fig:amd-specjbb-memory}
]

\clearpage

\end{document}